# Creep stability of the proposed AIDA mission target 65803 Didymos: I. Discrete cohesionless granular physics model

Yun Zhang[a,b], Derek C. Richardson[a], Olivier S. Barnouin[c], Clara Maurel[d], Patrick Michel[e], Stephen R. Schwartz[e], Ronald-Louis Ballouz[a], Lance A. M. Benner[f], Shantanu P. Naidu[f], Junfeng Li[b]

[a] *Department of Astronomy, University of Maryland, College Park, MD 20740-2421, United States*
[b] *School of Aerospace Engineering, Tsinghua University, Beijing 100084, China*
[c] *The Johns Hopkins University, Applied Physics Laboratory, USA*
[d] *Institut Supérieur de l'Aéronautique et de l'Espace-Supaéro, France*
[e] *Lagrange Laboratory, University of Nice Sophia Antipolis, CNRS, Observatoire de la Côte d'Azur, France*
[f] *Jet Propulsion Laboratory, California Institute of Technology, USA*

**Abstract** As the target of the proposed Asteroid Impact & Deflection Assessment (AIDA) mission, the near-Earth binary asteroid 65803 Didymos represents a special class of binary asteroids, those whose primaries are at risk of rotational disruption. To gain a better understanding of these binary systems and to support the AIDA mission, this paper investigates the creep stability of the Didymos primary by representing it as a cohesionless self-gravitating granular aggregate subject to rotational acceleration. To achieve this goal, a soft-sphere discrete element model (SSDEM) capable of simulating granular systems in quasi-static states is implemented and a quasi-static spin-up procedure is carried out. We devise three critical spin limits for the simulated aggregates to indicate their critical states triggered by reshaping and surface shedding, internal structural deformation, and shear failure, respectively. The failure condition and mode, and shear strength of an aggregate can all be inferred from the three critical spin limits. The effects of arrangement and size distribution of constituent particles, bulk density, spin-up path, and interparticle friction are numerically explored. The results show that the shear strength of a spinning self-gravitating aggregate depends strongly on both its internal configuration and material parameters, while its failure mode and mechanism are mainly affected by its internal configuration. Additionally, this study provides some constraints on the possible physical properties of the Didymos primary based on observational data and proposes a plausible formation mechanism for this binary system. With a bulk density consistent with observational uncertainty and close to the maximum density allowed for the asteroid, the Didymos primary in certain configurations can remain geo-statically stable without including cohesion.

# 1. Introduction

A significant number of binary asteroid systems are observed to have primaries with spin rates very close to the critical state at which the equatorial material would become gravitationally unbound (Pravec and Harris, 2007). In addition, the shapes of these primaries are almost spheroidal and some appear to have an equatorial ridge based on radar observations (e.g., Ostro et al., 2006). An explanation for these remarkable characteristics is provided by the theory of YORP-induced rotational disruption of rubble-pile asteroids (Bottke et al., 2002; Scheeres, 2007; Pravec et al., 2010; Walsh and Jacobson, 2015). The YORP effect can impart rotational torque on small bodies due to absorption and reemission of thermal radiation (Rubincam, 2000), which has been observed to change the obliquity and spin rates of several asteroids (e.g., Vokrouhlický et al., 2003; Taylor et al., 2007; Kaasalainen et al., 2007). Driven by the YORP thermal effect, asteroids with the structure of cohesionless self-gravitational aggregates could be rotationally accelerated to their critical spin limits and shed mass to form secondaries. Using numerical simulations of the mass shedding process and subsequent reaccumulation of debris in orbit around the primary, Walsh et al. (2008) re-created many of the properties of the observed binary systems. By introducing cohesive forces into similar rubble-pile numerical models, Sánchez and Scheeres (2016) found spinning self-gravitating aggregates could be directly split into two parts and form binary systems without gravitational re-accumulation.

Apart from the binary formation mechanism, another interesting problem related to binary systems is the spin rates of the primaries. How can some primaries stay at such high spin rates? Are these primaries still in the process of shedding mass? What are the plausible internal structures and physical parameters of these primaries? Given that a rubble-pile structure is considered to be an appropriate model for most small asteroids (Richardson et al., 2002), investigating the critical spin limits of rubble-pile asteroids can shed some light on these questions. An important motivation of this work is to gain insight into the dependence of the spin limits of self-gravitating aggregates on several physical properties.

Based on a comparison of the gravity and centrifugal force at the equator of a spinning spherical body, Harris (1996) derived the spin period limit for rubble-pile asteroids, e.g., 2.1 h for a body with a bulk density of ~2.5 g/cc. Holsapple (2001) pointed out that the critical spin rate is constrained by the shear strength of an asteroid, with lower strength resulting in a narrower region of permissible equilibrium spin states. Based on elastic-plastic continuum theory for solid materials, analytic expressions for the permissible spin rates as functions of the Mohr-Coulomb friction angle, $\phi$, for ellipsoidal bodies are presented in the work of Holsapple (2001, 2004). Given the discrete nature of rubble-pile asteroids, Richardson et al. (2005) used a Hard-Sphere Discrete Element Method (HSDEM) to numerically explore the shape and spin limits for ellipsoidal bodies, and found the results are consistent with the theory of Holsapple (2001). However, the rotational disruption mass-loss behavior of rubble-pile models obtained in HSDEM simulations (Walsh et al., 2008, 2012) is different from the analytical theory of Holsapple (2010), which showed that rubble piles will reshape to prevent mass loss. By using a Soft-Sphere Discrete Element Method (SSDEM), Sánchez and Scheeres (2012) suggested that the difference in the two failure modes arises because HSDEM cannot appropriately take the physical friction between particles into consideration. By turning on particle-particle friction in SSDEM, they obtained similar results to Holsapple (2010). The finite element modeling by Hirabayashi and Scheeres (2015) also confirmed that a homogeneous spherical body should experience failure in its central region at high spin rates, so landslides and mass shedding are unlikely to happen. However, the surface-shedding failure behavior is possible for spherical bodies with heterogeneous internal structure (Hirabayashi et al., 2015). In addition, for rubble-pile bodies, the arrangement and size distribution of constituent particles may also have impact on the spin limits and the failure behaviors. Here, we explicitly explore the effect of the internal configuration of rubble-pile bodies from a granular mechanics perspective. Although the rotational failure modes are different for different models, the critical spin rate shows clear dependence on the value of the material friction angle, $\phi$, and the bulk density, $\rho_B$, of rubble piles in all of these studies, which are also investigated in this work.

Instead of using spherical or ellipsoidal bodies as substitutes for asteroids, we focus on a real object, the primary of near-Earth binary asteroid 65803 Didymos (1996 GT). Didymos is the target of the proposed Asteroid Impact & Deflection Assessment (AIDA) mission (planned for fall 2022 encounter), which combines a kinetic impactor experiment (Cheng et al., 2016) and an orbiter to characterize the target and monitor the deflection results from the impact (Michel et al., 2016). AIDA will be the first kinetic impactor experiment carried out at real asteroid scales, and will return fundamental information on the physical and dynamical properties of a binary asteroid system and its mechanical response to an impact event. The spin period of the Didymos primary, ~2.26 h, may be close to its spin limit, which is the most well-constrained information about this asteroid system (Table 1). The present work is supporting the AIDA mission proposal by investigating the plausible physical properties of the primary based on current knowledge.

**Table 1** Didymos system basic properties.

| | |
|---|---|
| Primary diameter, $D_P$ | 0.780 km ± 10% |
| Secondary diameter, $D_S$ | 0.163 ± 0.018 km |
| Total system mass, $M_{TOT}$ | $(5.278 \pm 0.04) \times 10^{11}$ kg |
| Component bulk density, $\rho_B$ | 2100 kg/m$^3$ ± 30% |
| Primary rotation period, $T$ | 2.2600 ± 0.0001 h |

In this study, assuming the primary has a cohesionless self-gravitational rubble-pile structure, the effects of internal configuration, bulk density, and material friction angle on the creep stability are investigated using a high-efficiency SSDEM code. We define the terminology of creep stability used in this study as the ability of a rubble pile to hold its current shape at a given spin state without any creep deformation (the creep motion usually happens in sheared granular systems and causes instability; see Section 4.3 for details about creep behavior of granular materials).[1] Correspondingly, a rubble pile is called being geo-statically stable if no creep deformation occurs. Section 2 details the numerical method and the simulation parameters. To link critical spin limits with the failure criteria of granular materials, stress analyses of rubble-pile structures from a granular mechanics perspective are necessary. The stress analysis method and three devised definitions of critical spin limits are introduced in Section 3. With the goal of finding out possible physical properties for the Didymos primary that can allow it to keep its shape at the observed spin rate, Section 4 presents our numerical efforts to explore the effects that can improve the creep stability of rubble piles, and Section 5 summarizes the implications of these simulation results.

---

[1] Specifically, the creep stability of rubble-pile bodies defined in this study may be different from the structural stability of rubble-pile bodies subject to small perturbations, as studied by, e.g., Holsapple (2004) and Sharma (2013). We do not explicitly investigate the effect of perturbations in this study. The connection between these two definitions of stability is left for future study.

Readers interested primarily in the results can skip most of Section 2, but should read Section 2.3.2 to learn about the six tested rubble-pile configurations, summarized in Table 2. The discussion of critical spin limits in Section 3 is quite technical; we consider the third spin limit, denoted by $T_{c,3}$, to be the most representative of rubble-pile structural failure, so it can be adopted as the quantity to consider in the Section 4 results. Section 4.1 presents each rubble-pile configuration in turn using a parallel structure, so it is possible to just skip to the configuration of interest. Sections 4.2–4.4 summarize the effects of bulk density, spin-up path (the rotational acceleration profile of the rubble pile used in this study), and friction, respectively, and are largely self-contained. The discussion in Section 5 is recommended for all readers.

**2. Methodology**

2.1 SSDEM implemented in *pkdgrav*

In the present work, the Didymos primary is modeled as a self-gravitational aggregate of smaller, indestructible, but compressible spheres interacting with one another. Within a parallel *N*-body gravity tree code framework, *pkdgrav* (Richardson et al., 2000; Stadel, 2001), a soft-sphere discrete element method (SSDEM) is used for computing particle contact forces (Schwartz et al., 2012). The force acting on each particle is given by

$$\mathbf{F}_i = \sum_{j=1, j\neq i}^{N} \mathbf{F}_{ij}^{(g)} + \sum_{j=1}^{N_C} \mathbf{F}_{ij}^{(c)} , \qquad (1)$$

where $N$ is the total number of particles, $\mathbf{F}_{ij}^{(g)}$ and $\mathbf{F}_{ij}^{(c)}$ are the gravitational pull and contact force (if it exists) of particle *j* on particle *i*, and $N_c$ is the coordination number of particle *i* (that is, the number of contacts at that instant). When the contact surfaces are not frictionless, the particles in contact can also exhibit resistance to the relative tangential motion of their surfaces, which will impose a torque on these particles. The motion of a particle can be obtained by integrating the force and torque.

In *pkdgrav*'s soft-sphere implementation, a linear spring-dashpot model is used to describe the normal contact force $\mathbf{F}_N$ and the tangential stick-slip force $\mathbf{F}_S$ (Cundall and Strack, 1979). In brief, the contact forces between two particles are given by

$$\mathbf{F}_N = -k_N x \, \hat{\mathbf{n}} + C_N \mathbf{u}_n,$$
$$\mathbf{F}_S = \min\left(k_S \boldsymbol{\delta}_S + C_S \mathbf{u}_t, \mu_S |\mathbf{F}_N| \boldsymbol{\delta}_S / |\boldsymbol{\delta}_S|\right), \quad {}^2 \qquad (2)$$

which depend on the spring constants, $k_N$ and $k_S$, and the plastic damping parameters, $C_N$ and $C_S$ (which are related to the normal and tangential coefficients of restitution, $\varepsilon_n$ and $\varepsilon_t$). The variable $x$ is the mutual compression between two particles, and $\boldsymbol{\delta}_S$ is the sliding displacement from the equilibrium contact point. The unit vector $\hat{\mathbf{n}}$ gives the direction from one particle's center to its neighbor's center. The dashpot force is linearly proportional to the normal relative velocity and tangential relative velocity $\mathbf{u}_n$ and $\mathbf{u}_t$. The parameter $\mu_S$ is the interparticle friction coefficient.[3] A plastic twisting and rolling friction model is also implemented in *pkdgrav* (Schwartz et al., 2012), and we introduce a version more suited to the quasi-static regime in Section 2.2. A second-order leapfrog method is applied to solve the equations of motion. The numerical approach has been validated through comparison with laboratory experiments (e.g., Schwartz et al., 2014) and has been successfully used to study various behaviors of granular systems, i.e., the Brazil nut effect (Matsumura et al., 2014; Maurel et al., 2017), avalanche dynamics (Yu et al., 2014), and collisions between rubble-pile asteroids at low speeds (Ballouz et al., 2014, 2015; Zhang et al., 2015).

2.2 Rotational resistance model for quasi-static states

In addition to the normal and tangential forces described above, particles in contact can also exhibit resistance to rotational motions. Previous studies (e.g., Iwashita and Oda, 1998; Mohamed and Gutierrez, 2010) indicate that introducing rotational resistances into an SSDEM model can narrow the gap between the numerical predictions and laboratory experimental results. A significant number of rotational resistance models have been pro-posed with the aim of simulating different granular system states, i.e., the dynamic flow state, the quasi-static state, and the mixed condition where the two states coexist—see Ai et al. (2011) for detailed discussion. The original plastic twisting and rolling friction model implemented in *pkdgrav* (Schwartz et al., 2012) is best suited for the dynamic flow state, as outlined by Ai et al. (2011). Given that the YORP spin-up/spin-down timescale for kilometer-sized NEAs is estimated to be a few $10^4$ to $10^6$ years (Rubincam, 2000), a rubble pile undergoing substantial spin-state changes always stays in a quasi-static state before structural failure occurs. In order to capture this quasi-static behavior, an elastic-plastic spring-dashpot rotational resistance model is used in the present work.

---

[2] The min() function means that the vector quantity of least magnitude is the one chosen.
[3] Although the dynamic friction coefficient is generally smaller than the static friction coefficient for rocks, the difference is not significant at low sliding speeds (Heslot et al., 1994) and does not lead to qualitatively different behavior in SSDEM simulations (Luding, 2008). In this study, $\mu_S$ is taken as one constant for both static and dynamic regimes.

Similar to the normal and tangential forces, $\mathbf{F}_N$ and $\mathbf{F}_S$, the rotational resistance can be decomposed into twisting and rolling resistances. Figure 1 presents the directions of the forces and torques acting on particle *i* generated by the contact with particle *j*, where the twisting and rolling torques, $\mathbf{M}_T$ and $\mathbf{M}_R$, are introduced in the following sub-sections.

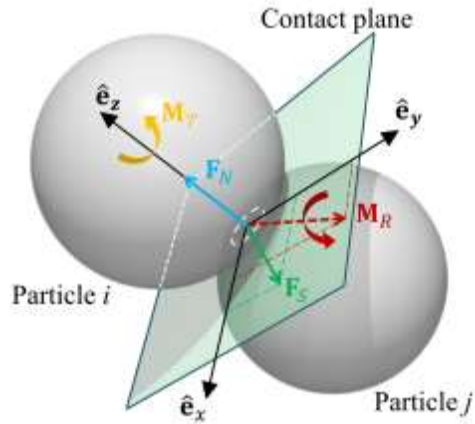

**Figure 1** Schematic of transmitted forces and torques at a contact in a local Cartesian coordinate system, where $\hat{\mathbf{e}}_x$ and $\hat{\mathbf{e}}_y$ are in the contact plane and $\hat{\mathbf{e}}_z$ is in the normal direction according to the right-hand rule. The dashed circle on the contact plane denotes the contact area of particle *i* and *j*.

2.2.1 Twisting resistance

The twisting resistance mostly arises from the slip and friction at the contact region due to a difference in the rotation rate of the particles in a direction along the normal vector, $\hat{\mathbf{n}}$. When two particles *i* and *j* are in contact, the twisting motion of particle *j* relative to particle *i* can be expressed in terms of their angular velocities $\boldsymbol{\omega}_i$ and $\boldsymbol{\omega}_j$ as

$$\boldsymbol{\omega}_T = \left[(\boldsymbol{\omega}_j - \boldsymbol{\omega}_i) \cdot \hat{\mathbf{n}}\right]\hat{\mathbf{n}}, \tag{3}$$

which results in a torque $\mathbf{M}_T$ oriented parallel to $\hat{\mathbf{n}}$ acting on particle $i$. The twisting spring-dashpot-slider model has a form similar to the tangential force, giving the resistance torque in the normal direction as

$$\mathbf{M}_T = \begin{cases} k_T \boldsymbol{\delta}_T + C_T \boldsymbol{\omega}_T, & \text{if } |k_T \boldsymbol{\delta}_T| < M_{T,\max} \\ M_{T,\max} \boldsymbol{\delta}_T / |\boldsymbol{\delta}_T|, & \text{if } |k_T \boldsymbol{\delta}_T| \geq M_{T,\max} \end{cases}, \tag{4}$$

where $k_T$ and $C_T$ are the twisting stiffness and viscous damping coefficients, respectively. The twisting angular displacement is given as

$$\boldsymbol{\delta}_T = \int_{t_0}^{t} \boldsymbol{\omega}_T(\tau) d\tau + \boldsymbol{\delta}_{T0}, \tag{5}$$

where the integral is over the duration of the contact before the critical twisting torque $M_{T,\max}$ is exceeded (i.e., the interval over which static twisting friction is acting). The initial twisting extension $\boldsymbol{\delta}_{T0}$ is zero when particles first penetrate, but can be nonzero in the event of twisting. Notice that, as particle pair $i$ and $j$ move, not only will the contact point move, but the equilibrium contact point will also change in the reference frame of the whole system. We account for this motion at every step by transforming $\boldsymbol{\delta}_T$ according to the change in $\hat{\mathbf{n}}$ over the previous step (in a way analogous to Schwartz et al., 2012, for the tangential displacement vector). We use a similar procedure for the rolling resistance.

2.2.2 Rolling resistance

The physical sources of rolling resistance are various, such as the slip and friction on the contact surface (Tabor, 1955), the viscous hysteresis (May et al., 1959), shape effect (Wensrich and Katterfeld, 2012), and surface adhesion (Li et al., 2011). Generally, rolling torque imposed on a particle is set in such a direction as to resist the rolling motion relative to the other particle in contact. It is hard to clearly distinguish the rolling motion from the sliding and rigid-body rotation of particles in continuous contact, especially when they move in different directions. There has been considerable debate about how to appropriately define rolling and sliding (Wang et al., 2015).

Following the definition of rolling displacement in Iwashita and Oda (1998) and Bagi and Kuhn (2004), we define the rolling velocity that characterizes rolling motion during collision of two spherical particles as

$$\mathbf{u}_R = \frac{l_i l_j}{l_i + l_j}(\boldsymbol{\omega}_j - \boldsymbol{\omega}_i) \times \hat{\mathbf{n}} + \frac{1}{2}\frac{l_j - l_i}{l_j + l_i}\mathbf{u}_t, \tag{6}$$

where $l_i$ and $l_j$ are the contact lever arms from the center of particle $i$ and $j$ to the effective point of contact, respectively. We refer the reader to Iwashita and Oda (1998) and Bagi and Kuhn (2004) for a detailed derivation of the rolling velocity, in which they applied the particle radii in this formula instead of the lever arms. Since the actual relative motion at the contact point is related to the lever arm, the contact lever arm is preferred in our implementation. This treatment does not make a significant difference so long as the maximum overlap during the simulation is small enough (we keep the maximum overlap smaller than 1% of the smallest particle radius in our simulations). If the spheres in contact are of equal size, Eq. (6) reduces to the simple expression

$$\mathbf{u}_R = \frac{l_i}{2}(\boldsymbol{\omega}_j - \boldsymbol{\omega}_i) \times \hat{\mathbf{n}}. \tag{7}$$

Assuming the lever arm of the rolling motion is $l_i l_j / (l_i + l_j)$, the relative rolling rate is

$$\boldsymbol{\omega}_R = \frac{l_i + l_j}{l_i l_j}\hat{\mathbf{n}} \times \mathbf{u}_R = \boldsymbol{\omega}_j - \boldsymbol{\omega}_i - \left[(\boldsymbol{\omega}_j - \boldsymbol{\omega}_i) \cdot \hat{\mathbf{n}}\right]\hat{\mathbf{n}} + \frac{1}{2}\frac{l_j - l_i}{l_i l_j}\hat{\mathbf{n}} \times \mathbf{u}_t. \tag{8}$$

Similar to the twisting interaction, the rolling spring-dashpot-slider model is given as

$$\mathbf{M}_R = \begin{cases} k_R \boldsymbol{\delta}_R + C_R \boldsymbol{\omega}_R, & \text{if } |k_R \boldsymbol{\delta}_R| < M_{R,\max} \\ M_{R,\max} \boldsymbol{\delta}_R / |\boldsymbol{\delta}_R|, & \text{if } |k_R \boldsymbol{\delta}_R| \geq M_{R,\max} \end{cases}, \tag{9}$$

where $k_R$ and $C_R$ are the rolling stiffness and viscous damping coefficients, respectively. The rolling angular displacement is given as

$$\boldsymbol{\delta}_R = \int_{t_0}^{t} \boldsymbol{\omega}_R(\tau) d\tau + \boldsymbol{\delta}_{R0}, \tag{10}$$

where the integral is over the duration of the contact before the critical rolling torque $M_{R,\max}$ is exceeded (i.e., the interval over which static rolling friction is acting), and $\boldsymbol{\delta}_{R0}$ is the initial rolling extension (which is zero when particles first penetrate). Similar to the twisting angular displacement, the rolling angular displacement $\boldsymbol{\delta}_R$ is adjusted based on the rotation of the contact plane around the vector $\hat{\mathbf{n}}$ at every time step.

2.2.3 Model parameters

The framework of the spring-dashpot-slider model given by Eqs. (4) and (9) has been used by a number of investigators and there are several different methods on how the model parameters should be set for 2D and 3D systems (e.g., Iwashita and Oda, 1998; Jiang et al., 2005; Mohamed and Gutierrez, 2010; Ai et al., 2011). There are six parameters that need to be decided on for the twisting and rolling spring-dashpot-slider models, i.e., ($k_T$, $C_T$, $M_{T,\max}$) and ($k_R$, $C_R$, $M_{R,\max}$). Recently, Jiang et al. (2015) derived physically based model parameters for the 3D elastic-plastic spring-dashpot type contact model. The contact force over a contact area of two particles can be represented by an infinite number of continuously distributed force elements. Based on ideal distributions of the normal and tangential force elements on an assumed circular contact area, they expressed the contact behavior in the twisting and rolling directions as integrations of the normal and tangential force elements. Additionally, they regarded the shape effect as one of the primary physical mechanisms for the rotational resistance, which makes the model applicable to simulated systems of non-spherical particles.

By defining a shape parameter $\beta$ to represent a statistical measure of real particle shape, the twisting/rolling stiffness and damping coefficients can be expressed as

$$k_T = 2k_S (\beta R)^2, \ C_T = 2C_S (\beta R)^2,$$
$$k_R = k_N (\beta R)^2, \ C_R = C_N (\beta R)^2, \tag{11}$$

where the effective radius $R = r_i r_j /( r_i + r_j)$, and $r_i$ and $r_j$ are the radii of the corresponding spherical particles. When the relative twisting/rolling rotational motion exceeds a critical twisting/rolling rotation, the force elements on the edge of the contact area begin to break, and the peak twisting/rolling torque is given as

$$M_{T,\max} = \mu_T \beta R \mu_S |\mathbf{F}_N|,$$
$$M_{R,\max} = \mu_R \beta R |\mathbf{F}_N|. \tag{12}$$

The static friction coefficients for twisting and rolling, $\mu_T$ and $\mu_R$, describe the hardness of the particle material. The determination of these parameters is based on the physical origin of plasticity and we refer the reader to Jiang et al. (2015) for details. See Section 2.3.3 for our adopted values.

The analyses from conventional triaxial and plane-strain compression tests conducted by Jiang et al. (2015) show that using these definitions of the material parameters can capture the quasi-static mechanical behavior of granular matter, and the shape parameter $\beta$ can well reflect the deviation of particle shape from spherical. Furthermore, the model needs only three additional physically based parameters, ($\mu_T$, $\mu_R$, $\beta$), which improves the efficiency of the parameter space exploration of SSDEM. The specific parameters used in the present work are introduced in Section 2.3.3.

2.3 Simulation parameters

2.3.1 Dynamical and physical properties of 65803 Didymos

Table 1 gives basic properties on the current dynamical and rotational state of Didymos determined from Earth-based radar and optical telescopic observations (Richardson et al., 2016). Under the assumption that the two components of the binary have the same bulk density, the mass of the primary can be estimated based on the diameter ratio, $M_P \approx 5.23 \times 10^{11}$ kg. Figure 2 shows the surface slopes[4] on the primary with the current shape model from radar observations. The effective resolution of the shape model is about 50 m. We use this model to construct the configurations of our rubble-pile models. It is assumed that the primary is a principal-axis rotator, in which case the rotation pole is along the maximum-moment principal axis of inertia, i.e., the $z$ axis shown in Fig. 2.

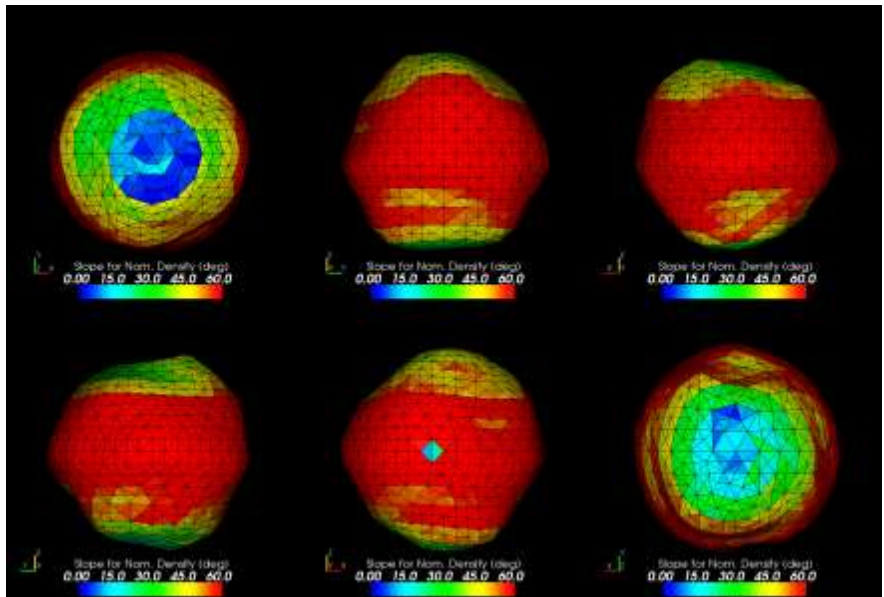

**Figure 2** Surface slopes on the Didymos primary from various perspectives based on the latest radar shape model and the parameters in Table 1. Note that slopes greater than 50° are well above any angle of repose of cohesionless terrestrial sands.

2.3.2 Rubble-pile models

---

[4] The surface slope is the angle between the normal of the facet and the effective local gravity vector.

To investigate the effect of internal configuration and particle size distribution on the mechanical behavior of a Didymos-analog rubble pile during spin-up, we explore six possible configurations in this study: 1. a hexagonal closest packing (HCP) model; 2. a monodisperse random close packing (RCP) model; 3. a monodisperse close packing model with a 200-m-radius HCP core and a random packing shell (henceforth, RCPC); 4. & 5. two dense polydisperse packing models with different size distributions (henceforth, PP1 & PP2); and 6. a dense polydisperse packing model with a 200-m-radius higher-density core (henceforth, PPC), where the grain density of particles in the core is 1.5 times greater than that of particles in the exterior. Except for the last case, the grain density distributions of these models are uniform. Although the grains of an actual asteroid are likely distributed in a random way, the HCP structure can serve as a material of higher shear resistance and improve our understanding of how the internal structure influences the failure mode of a spinning aggregate (Walsh et al., 2012).

Most of the aggregates are constructed using a two-step procedure. First, a rubble pile is created by placing spheres with a pre-defined size distribution randomly in a spherical cloud and allowing the cloud to gravitationally collapse with highly inelastic collisions. To increase the filling fraction, the gravitationally accumulated packing is subject to a vertical oscillation in a sinusoidal manner within a solid spherical boundary in a uniform gravity field (we use 1g). After vibration, the granular assembly is allowed to settle down under its own self-gravity with no external field. Frictionless material parameters are used for the above procedure to minimize the void space between particles as well as to avoid the "Brazil-nut effect" in the process of vibration (so the obtained rubble pile is macroscopically homogeneous; Tancredi et al., 2012; Matsumura et al., 2014; Maurel et al., 2017). Although the Brazil-nut effect may cause the migration of grains in rubble-pile bodies (Asphaug et al., 2001; Perera et al., 2016), we assume these complicated processes have not adversely affected the morphology of our models. Investigation of effects of particle redistribution due to seismic quakes or impacts on the spin-up process will be left for future studies. The HCP structure can be directly constructed without the above procedure. The model with HCP core can be obtained from a similar procedure combining the HCP structure with a random granular assembly. Second, the shape model of the Didymos primary is used to carve out the figure of the granular assembly for all cases.

Table 2 summarizes the basic properties of these rubble-pile models. The differential size distributions of the polydisperse models are assumed to follow a power law with an exponent of –3, as observed at the surface of several bodies and consistent with theoretical expectations (e.g., Michikami et al., 2010). Given that the resolution of the shape model is ~50 m, aggregates composed of particles of around 10-m radius are sufficient to match the shape.

**Table 2** Basic properties of rubble-pile models with the nominal bulk density. Here $N$ is the number of particles in the model. Particle radii in polydisperse models follow a differential power law of exponent –3 over the listed range. The PPC model has two densities listed, the first for the exterior shell and the second for the interior core.

| Rubble-pile Model | $N$ | Particle radius (m) | Grain density (g/cc) |
|---|---|---|---|
| HCP | 42062 | 10 | 2.97 |
| RCP | 36523 | 10 | 3.42 |
| RCPC | 37081 | 10 | 3.37 |
| PP1 | 47761 | 4–32 | 3.01 |
| PP2 | 95637 | 4–16 | 3.20 |
| PPC | 47761 | 4–32 | (2.80, 4.20) |

2.3.3 Material parameters—modeling different angles of friction

Previous studies show that the spin limits of rotating asteroids inherently depend on the angle of friction, which represents the slope that specifies shear strength under a given pressure (e.g., Holsapple, 2004; Sharma et al., 2009; Sánchez and Scheeres, 2016). In our SSDEM model, there are five parameters, ($k_S$, $\mu_S$, $\mu_T$, $\mu_R$, $\beta$), that give rise to shear strength. To keep normal and tangential oscillation frequencies equal to each other, the tangential spring constant $k_S$ is set to $(2/7)k_N$ (see Section 2.6 in Schwartz et al., 2012 for details). As suggested by Jiang et al. (2015), the static friction coefficients for twisting and rolling, $\mu_T$ and $\mu_R$, are set to 1.3 and 1.05, respectively, corresponding to sand particles of medium hardness.[5] The numerical experiments carried out by Jiang et al. (2015) indicate that higher shear strength can be achieved by using a higher friction coefficient, $\mu_S$, and more angular particles, i.e., a larger value of $\beta$. We use a set of parameters, $\mu_S =$ 1.0, $\beta = 0.5$, as the nominal material parameters, for which the estimated friction angle for a spinning rubble pile is within the typical range of terrestrial sands (25°–45°) observed in laboratory tests (see Table 3). The effect of $\mu_S$ and $\beta$ is also investigated in this study (see Section 4.4).

---

[5] For infinitely strong material, $\mu_T$ and $\mu_R$ would be 4/3 and 2.0, respectively.

The normal spring constant $k_N$ and the timestep $\Delta t$ are set to ensure that particle overlaps for each entire simulation do not exceed 1% of the minimum particle radius (see Schwartz et al., 2012 for details). The coefficients of restitution, $\varepsilon_n$ and $\varepsilon_t$, are set to 0.55 so that the granular system is subject to sufficient damping; these values are also reasonable for terrestrial rocks (Chau et al., 2002).

2.3.4 Quasi-static spin up

To precisely capture the mechanical behavior of rubble piles with varying spin rate, we implemented a spin-up procedure that sets the body's spin period $T$ to a prescribed value as a function of time $t$. For our experiments, we used a spin-up path like the one shown in Fig. 3. In the beginning of the simulation, the body settles down at a slow spin state $T_{start}$. The spin-up path is divided into two sub-paths with different spin-up rates. First, the body is linearly spun up from $T_{start}$ to $T_{inter}$ in a relatively rapid way, where $T_{inter}$ is sufficiently far from the critical value to ensure the body always remains stable. Then, the spin period is decreased from $T_{inter}$ to $T_{final}$ in a slower manner, including several spin-up stages and settle-down stages. A more realistic acceleration method would need to consider the coupled effect of the shape and the YORP torque, as discussed in Cotto-Figueroa et al. (2015). Within the scope of this work, we do not go further in analyzing the precise spin-up behavior caused by the YORP effect. The spin period of the simulated body is strictly constrained to the spin-up path shown in Fig. 3 before global disruption occurs.[6]

---

[6] The global disruption event is very obvious to distinguish, where the values of the packing efficiency and the axis ratios rapidly drop. When the global disruption is confirmed, we stop maintaining the spin period of the simulated body, and let the body evolve freely under its own gravity. In general, since the structure is disrupted, the spin period of the body will abruptly increase after we set it free. For example, as shown in Fig. 5, the spin period $T$ keeps decreasing (the top frame) until the packing efficiency and axis ratios abruptly drop (the lower-middle frame).

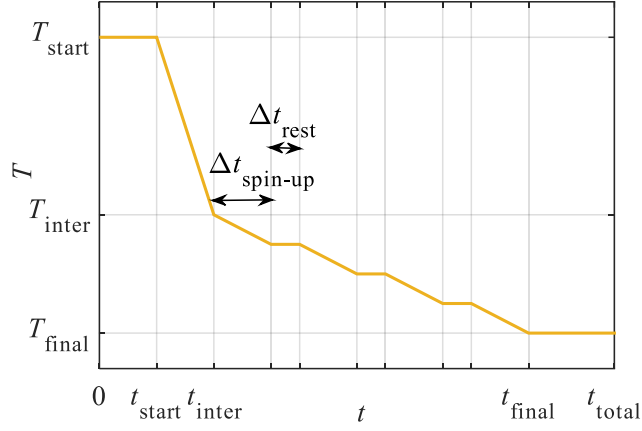

**Figure 3** Illustration of the spin-up path (rotation period $T$ as a function of time $t$) for the spin-up tests. $t_{\text{start}}$ is the duration of the interval from the start of the simulation, when the rubble pile has a spin period of $T_{\text{start}}$. $t_{\text{inter}}$ denotes the beginning of the slower spin-up, when the rubble-pile has a period of $T_{\text{inter}}$. The spin-up process is finished at $t_{\text{final}}$, and the rubble pile is forced to stay at the spin period of $T_{\text{final}}$ after that.

For our nominal case (i.e., path No. 1 in Table 5), we use about 7000 s to go from $T_{\text{start}}$ to $T_{\text{inter}}$, which are set to 5 h and 3 h, respectively, and another 320000 s to achieve the final spin state, $T_{\text{final}} = 2.26$ h. The bodies are allowed to settle down five times for about 7000 s each time ($\Delta t_{\text{rest}}$ in Fig. 3) in between spin-up stages of about 47500 s ($\Delta t_{\text{spin-up}}$ in Fig. 3). The parameters governing the process were chosen so that the simulations remain computationally expedient while the aggregates stay in quasi-equilibrium states before structural failure occurs. Given that the dynamical time for a rubble pile is $1/\sqrt{G\rho_B}$, where $G$ is the gravitational constant and $\rho_B$ is the body's bulk density, i.e., approximately 2650 s for the Didymos primary, the time scale of this spin-up process is sufficient to maintain quasi-equilibrium. Nevertheless, this spin-up time is much shorter than a true YORP timescale. To test the effect of spin-up path on the mechanical behavior of spinning aggregates, simulations were run using a few different spin-up paths (Section 4.3).

## 3. Critical spin limits

In order to obtain the critical spin rate/period, we need to define a measurement for determining whether the state of a self-gravitating aggregate is below its spin limit. Previous discrete-element simulations on spinning self-gravitating aggregates often take sub-stantial changes in the shape or mass-loss behavior during the spin-up process as a sign of structural failure, and assign the spin rate at the time this occurs to the critical spin rate (e.g., Walsh et al., 2012; Sánchez and Scheeres, 2012). However, since the maximum shear stress is often located at a body's center at a rapid spin state, its internal structure may fail first without a clear surface expression (Hirabayashi and Scheeres, 2015), which implies a lower critical spin rate.

In this study, we devise three different methods to characterize the spin limit, which are detailed in the following sub-sections. The shape, packing efficiency, and stress states of the body are monitored throughout the spin-up process for each simulation (e.g., see Fig. 5). By defining three critical spin limits based on these variables, the critical states triggered by variations in external shape and surface shedding (Section 3.1), internal structural deformation (Section 3.2), and shear yielding (Section 3.3) can all be captured.[7]

3.1 First critical spin limit: reshaping and surface shedding

The *first critical spin rate*, $\Omega_{c,1}$, and the *first critical spin period*, $T_{c,1}$, are designed to reflect the reshaping and surface shedding process—they are defined at the point when the change in one of the axis ratios, $b/a$ or $c/a$, exceeds 0.01 compared to the initial value. The axis lengths $a$, $b$, and $c$ are defined as the maximum extensions along the principal axes of inertia (from largest to smallest) of the whole granular system (including any particles that are shed from the surface). The tolerance of 0.01 is chosen because the overlap between particles in contact is less than 1% of the particles' mean radius and expansion above this value indicates considerable change in shape has occurred. The first critical spin limit therefore gives information about reshaping processes as well as surface-shedding behaviors.

3.2 Second critical spin limit: internal structural deformation

The internal deformation of a spinning rubble pile will cause subtle changes in its internal porosity, which can be measured by using a Voronoi tessellation of the rubble pile. This divides the space containing all particles into subdomains, one per particle, where each subdomain consists of all points closer to its particle than any other (Okabe et al., 2009). In 3D Euclidean space, a Voronoi cell is an irregular convex polyhedron, as illustrated in Fig. 4(a). We use the open-source library, *Voro++* (Rycroft, 2009), to build the Voronoi tessellation of our rubble piles. Figure 4(b) shows the Voronoi diagram for the Didymos primary, where the outer boundary for the tessellation is its radar shape model. The Voronoi tessellation expresses the discrete granular aggregate as an appropriate equivalent continuum, allowing us to conduct the stress analyses with conventional continuum methods for a granular medium (see Section 3.3.1).

---

[7] In this study, the term "surface shedding" is the process in which particles resting on the asteroid surface are observed to move on and be lifted from the surface, and the term "internal structural deformation" refers to the rearrangement process of constituent particles inside a rubble pile.

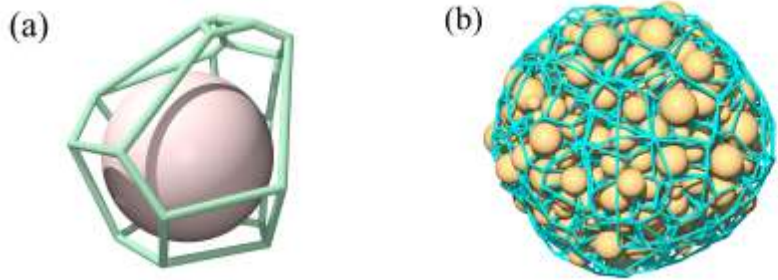

**Figure 4** The Voronoi tessellation for (a) a particle in a granular assembly and (b) a granular representation of the Didymos primary, as computed using the *Voro++* library. Note that some particles stick out at the surface since particle radii are ignored when filling the radar shape model. Given that the maximum particle radius (i.e., 32 m) used in this study does not exceed the resolution of the shape model (i.e., 50 m), this does not adversely affect the quality of the shape matching.

Voronoi tessellation can provide a consistent measure of the internal filling fraction of granular assemblies. For a given particle $i$ with a radius of $r_i$, its packing fraction in its own Voronoi cell is

$$\eta_i = 4\pi r_i^3 / (3V_i). \tag{13}$$

where $V_i$ is the volume of the corresponding Voronoi cell for the $i$-th particle. The average packing fraction of the entire aggregate with $N$ particles is then

$$\eta_B = \frac{1}{N}\sum_{i=1}^{N} \eta_i . \tag{14}$$

Note that the shapes of Voronoi cells for surface particles are constrained by the boundary condition of the tessellation (i.e., the shape model of the Didymos primary in this case), so they should not be used to characterize the void space between particles. In practice, to eliminate the effect of boundary conditions and capture the internal deformation of an aggregate, we only take internal particles into account to calculate the average packing efficiency.

The second critical spin limit is designed to reflect the internal deformation. Associated with the spin-up process, the aggregate will begin to exhibit dilation behavior and the packing efficiency will decrease significantly. The moment when the internal packing fraction, $\eta_B$, changes by more than 0.001 compared to the initial value corresponds to the *second critical spin rate*, $\Omega_{c,2}$, and the *second critical spin period*, $T_{c,2}$. Similar to the first critical spin limit, the tolerance of 0.001 is chosen because a relative displacement between two particles in contact of 1% of the particles' mean radius will cause a variation of ~0.001 in the packing efficiency (see Appendix).

3.3 Third critical spin limit: shear failure

The behavior of reshaping or internal deformation is just the mechanical response of granular materials to shear failure. Therefore, it is useful to analyze the stress states to check if the spinning aggregate is globally geo-statically stable.

3.3.1 Average stress in granular materials

Macroscopic stress variables in a discrete element assembly can be assessed by homogenization and averaging methods based on Representative Volume Elements (RVEs) with the help of Voronoi tessellation (Bagi, 1996; Luding, 2004). A RVE should contain a sufficient number of particles to eliminate fluctuations at the microscopic scale. On the other hand, a RVE should also be much smaller than the macroscopic dimensions of the whole computational domain if the desired macroscopic quantity to be measured is non-homogeneous. The size of a RVE is often taken to be about 12 times the radius of the typical particle for stress analyses (Masson and Martinez, 2000). We use a tree code to divide the granular aggregate into several similar-sized sub-assemblies, i.e., RVEs. Each RVE contains ~300 particles.

At the microscopic scale, the discrete particles in our granular systems interact with each other via contact forces and torques determined by the model described in Section 2.

Based on these microscopic variables, we apply a two-step averaging procedure to calculate the macroscopic stress tensor, following Bagi (1996). In the first step, from Cauchy's first law of motion, the Cauchy stress tensor for a single particle $i$ is given as the summation of the dyadic product of the branch vector $\mathbf{x}^{i,k}$ that connects the particle center with the contact point for each contact and the corresponding contact force $\mathbf{f}^{i,k}$,

$$\boldsymbol{\sigma}_i = \frac{1}{V_i} \sum_{k=1}^{N_{i,c}} \mathbf{x}^{i,k} \otimes \mathbf{f}^{i,k}, \tag{15}$$

where the $N_{i,c}$ is the coordination number (number of contacts) and $V_i$ is the volume of the corresponding Voronoi cell for the $i$-th particle. In the second step, by volume-weighted averaging over a RVE, the average stress tensor in the $j$-th RVE can be calculated as

$$\bar{\boldsymbol{\sigma}}_j^{\text{RVE}} = \frac{1}{V_{\text{RVE}}} \sum_{i=1}^{N_{\text{RVE}}} V_i \boldsymbol{\sigma}_i = \frac{1}{V_{\text{RVE}}} \sum_{i=1}^{N_{\text{RVE}}} \sum_{k=1}^{N_{i,c}} \mathbf{x}^{i,k} \otimes \mathbf{f}^{i,k}, \tag{16}$$

where $N_{\text{RVE}}$ is the total number of particles in this RVE. The volume for each RVE, $V_{\text{RVE}}$, is the summation over the volume of the Voronoi cell of every particle in each RVE. We will use the average stress tensor to compute the yield criterion for our rubble piles.

3.3.2 Failure criterion

In the rock and soil mechanics community, the Mohr-Coulomb yield criterion is one of the most common yield criteria for geophysical granular materials (Jaeger et al., 2009) that is also used in structure failure analyses for rubble-pile asteroids (e.g., Holsapple, 2004; Hirabayashi et al., 2015).[8] In the Mohr-Coulomb model, there is a maximum shear stress that a granular material can withstand, which in turn depends on the compressive normal stress $\sigma_n$,

$$\tau \leq Y - \sigma_n \tan \phi, \tag{17}$$

where $Y$ is the cohesive strength and $\phi$ is the angle of friction. With the help of Mohr's circle, Eq. (17) can be written as a function of the principal stress,

$$\sigma_1 - \sigma_3 + (\sigma_1 + \sigma_3) \sin \phi \leq 2Y \cos \phi, \tag{18}$$

---

[8] Note that the Drucker-Prager yield criterion (a smooth version of the Mohr-Coulomb yield criterion) is also used for studying rubble-pile asteroids, and may match simulations better in some cases (Sharma et al., 2009). However, for analyzing the theoretical upper spin limit of spinning asteroids, Holsapple (2007) and Sharma et al. (2009) both showed that these two criteria give essentially the same answers. To explicitly express the relation between the friction angle and the stress state variables, the Mohr-Coulomb yield criterion is preferred in this study.

where $\sigma_1$ and $\sigma_3$ are the maximum and minimum components of the principal stress, respectively. It should be noted that the values of the principal stress components are always negative since only the compressive contact interaction is considered in this study.[9] For cohesionless materials ($Y = 0$), there is a simple relation between the angle of friction and the principal stress components,

$$\sin\phi \geq -(\sigma_1 - \sigma_3)/(\sigma_1 + \sigma_3). \tag{19}$$

When the friction angle of the material is known, applying the stress-state variables to Eq. (19) can indicate if the structure has undergone plastic yielding or has already failed, in which case Eq. (19) is no longer satisfied. For the $j$-th RVE, the principal stress components, $\bar{\sigma}_{j,1}^{\mathrm{RVE}}$, $\bar{\sigma}_{j,2}^{\mathrm{RVE}}$, and $\bar{\sigma}_{j,3}^{\mathrm{RVE}}$ (from largest to smallest) can be evaluated though diagonalizing its average stress tensor $\bar{\boldsymbol{\sigma}}_{j}^{\mathrm{RVE}}$ (Eq. (16)).[10]

3.3.3 Angle of friction and third critical spin limit

The intrinsic material friction angle, $\phi$, needs to be determined for each simulated rubble pile in order to carry out the above analysis. Since the particular internal configuration of the discrete assembly affects its friction angle (see discussion in Section 5.2), the value needs to be obtained for each case simulated. Sánchez and Scheeres (2012, 2016) evaluated $\phi$ by solving the Drucker-Prager yield criterion when the first signs of global reshaping are visible. However, as discussed before, a rubble pile could fail locally before reshaping is observed in discrete element simulations. As the spin rate is continuously increased, the maximum shear stress in a body keeps rising to the point of structural yielding by shearing; this is when the rubble-pile structure becomes unstable. We define the spin rate at which the failure criterion, Eq. (19), is first violated, as the *third critical spin rate*, $\Omega_{c,3}$, and the corresponding period as the *third critical period*, $T_{c,3}$. In principle, this also gives the friction angle of the rubble pile, but there is a complication.

---

[9] In structural engineering, the compressive stress is expressed as a negative value, and the tensile stress is expressed as a positive value. We follow this convention in the calculation of stress variables.

[10] Note that the mean stress tensor $\bar{\boldsymbol{\sigma}}_{j}^{\mathrm{RVE}}$ defined by Eq. (16) can be non-symmetric in the presence of rotation resistances, which may cause trouble in the diagonalization. The study of Oda and Iwashita (2000) shows that the stress tensor is almost symmetric in quasi-static experiments and simulations of granular materials. In this work, by evaluating stress states of each RVE in an aggregate within the quasi-static process of spin-up, we confirmed that the mean stress tensors are nearly symmetric and the Mohr-Coulomb yield criterion can be applied without difficulty.

Since both sides of Eq. (19), i.e., the angle of friction and the stress-state variables at the critical state, are unknown, the critical state cannot be determined in a single spin-up simulation. To find the third critical spin limit in a systematic way, multiple spin-up simulations are carried out with different $T_{\text{final}}$ for each rubble-pile model to determine the critical state at which the rubble pile can be just geo-statically stable with the radar-derived shape without further creep deformation. The simulated rubble pile is forced to stay at the spin period of $T_{\text{final}}$ after the spin-up process, so the creep behavior in this body can be diagnosed when gradual increases in the shear stress or gradual decreases in the internal packing efficiency and coordination number are observed (see Fig. 22 for examples; also see the discussion in Section 4.3 for details). We continuously increase $T_{\text{final}}$ from $T_{\text{final}} = T_{c,2}$ (the second critical spin period is expected to be shorter than $T_{c,3}$ since the internal deformation occurs after shear yielding) in increments of 0.01 h until the critical state is achieved. Following this procedure, the values of $\Omega_{c,3}$ and $T_{c,3}$ are obtained. Then, the angle of friction for the corresponding rubble pile can be found from Eq. (19). However, due to the discrete nature of the material, the stress state will vary throughout the rubble-pile structure. Therefore, we define the angle of friction $\phi$ to be $\max_{j \in \text{RVE}}(\theta_j)$ at the critical state, where

$$\theta_j = \arcsin\left[\left(\bar{\sigma}_{j,3}^{\text{RVE}} - \bar{\sigma}_{j,1}^{\text{RVE}}\right) \Big/ \left(\bar{\sigma}_{j,1}^{\text{RVE}} + \bar{\sigma}_{j,3}^{\text{RVE}}\right)\right] \tag{20}$$

is what we call the "mobilized friction angle" for the *j*-th RVE at a particular time.[11] The rationale for this choice of $\phi$ is that the region subject to the highest shear stress is the most sensitive part to failure in a body (Hirabayashi et al., 2015). In practice, given the rapid oscillation behavior of the evolution curve of the maximum mobilized friction angle (i.e., $\max_{j \in \text{RVE}}(\theta_j)$ as a function of time; see Fig. 5 for an example), it is hard to accurately estimate the value of the friction angle at a certain state. Considering that the material shear resistance of a stable body should not be smaller than the shear stress acting on it, $\phi$ is taken to be the maximum in the interval $[T_{c,3} - 0.01 \text{ h}, T_{c,3} + 0.01 \text{ h}]$.

---

[11] We calculate the value of $\theta_j$ in a way analogous to the definition of the friction angle in the Mohr-Coulomb yield criterion (Eq. (19)), but it is not the friction angle that the simulated granular material may actually have. The word "mobilized" is used to distinguish it from the friction angle $\phi$.

## 4. Simulation Results

In order to find out possible physical properties of the Didymos primary that permit it to keep its shape at the observed spin rate, multiple spin-up simulations were carried out for all six rubble-pile models with various parameters. The following subsections present the critical spin limits and failure patterns determined for the nominal mass and shape, and detail what factors enhance the creep stability of rubble-pile bodies.

4.1 Nominal case

With the nominal material parameters and the nominal mass and shape model of the Didymos primary, the six rubble-pile models were spun up along the nominal spin-up path (see Section 2.3 for the definition of nominal parameters). Table 3 gives the initial packing efficiency of these models and summarizes the results of the simulations, which are detailed below.

**Table 3** Basic properties and spin limits of nominal rubble-pile models evaluated from our simulations. See Table 2 for the definitions of the models. Here $\eta_B^{initial}$ is the initial packing efficiency, $\phi$ is the friction angle as determined by the procedure given in Section 3.3.3, and $T_{c,1}$, $T_{c,2}$, and $T_{c,3}$ are the critical spin limits determined by shape change, internal deformation, and yield analysis, respectively.

| Rubble-pile Model | $\eta_B^{initial}$ | $\phi$ (deg) | $T_{c,1}$ (h) | $T_{c,2}$ (h) | $T_{c,3}$ (h) |
|---|---|---|---|---|---|
| HCP | 0.739 | 43 | 2.38 | 2.36 | 2.40 |
| RCP | 0.641 | 30 | 2.51 | 2.53 | 2.83 |
| RCPC | 0.695 | 32 | 2.46 | 2.46 | 2.76 |
| PP1 | 0.820 | 39 | 2.41 | 2.44 | 2.53 |
| PP2 | 0.750 | 38 | 2.41 | 2.44 | 2.54 |
| PPC | 0.820 | 39 | 2.46 | 2.41 | 2.49 |

4.1.1 Hexagonal-closest-packing configuration (HCP)

Figure 5 shows the evolution of the stress states of the HCP model, along with its spin period, packing fraction, axis ratios, mean coordination number (CN, i.e., mean number of contacts per particle), and the fraction of particles with only one ($N_c^1$) or no contacts ($N_c^0$) during the spin-up process. The stress states are represented by the maximum devi-atoric stress, the maximum pressure, and the maximum mobilized friction angle among all RVEs in the aggregate, i.e., $s_{\max} = \max\limits_{j \in \text{RVE}}(s_j)$, $p_{\max} = -\max\limits_{j \in \text{RVE}}(|p_j|)$ (the negative sign indicates compression), and $\theta_{\max} = \max\limits_{j \in \text{RVE}}(\theta_j)$, where

$$s_j = \left(\bar{\sigma}_{j,1}^{\text{RVE}} - \bar{\sigma}_{j,3}^{\text{RVE}}\right)/2 \text{, and} \tag{21a}$$

$$p_j = \left(\bar{\sigma}_{j,1}^{\text{RVE}} + \bar{\sigma}_{j,2}^{\text{RVE}} + \bar{\sigma}_{j,3}^{\text{RVE}}\right)/3, \tag{21b}$$

for the *j*-th RVE. Figure 6 illustrates the distribution of these stress-state variables and the force networks over a cross-section of the HCP structure.

The stress-state variables exhibit significant dependence on the spin period. At the rapid acceleration stage ($T_{\text{inter}} \leq T \leq 5$ h), $s_{\max}$ and $\theta_{\max}$ both occur at mid-latitudes on the surface, as shown in Fig. 6(1a) and (3a). Evidently, the surface region is the loosest part in the HCP structure (Walsh et al., 2012). The internal structure is subject to weak shear stress at such low spin rates, while the particles over mid-latitudes at the surface need higher tangential friction force to keep them stable because the highest surface slope of the shape occurs at the mid-latitude regions (Fig. 2), as indicated in Fig. 6(4a). As the spin rate is increased, the centrifugal forces parallel to the equatorial plane become larger, leading to a decrease in the magnitude of the internal pressure (see Fig. 5, upper-middle frame, and Fig. 6(2b), where the negative sign indicates compression). Meanwhile, since the forces along the vertical direction caused by gravity remain constant, the body experiences larger deviatoric stresses (Fig. 6(1b)) and the tangential friction forces between particles also grow (Fig. 6(4b)). When the spin rate increases up to a certain value, the force chains parallel to the horizontal plane near the equator begin to break (Fig. 6(4c)). Although some particles on the surface become rattlers with only one contact, the whole body can still remain stable when the rotational acceleration is stopped (see the fourth rest time interval in Fig. 5). The region where horizontal contacts are lost gradually expands to a larger area, resulting in a continuous decrease in the coordination number (the bottom frame in Fig. 5). Once the contact-breakage behavior propagates to the RVE with the current $\theta_{\max}$, $\theta_{\max}$ starts decreasing. The reason for the decrease is that when the topology of the force network changes, the remaining force chains attempt to maintain equilibrium of the system by adjusting themselves. By checking variations in the stress states in the RVE around this point in time, we find that the deviatoric stress no longer increases but decreases as the spin rate grows, while the pressure tends to remain constant, which leads to a smaller $\theta_{\max}$. Other RVEs also exhibit similar behavior when the force chains in the local region begin to break (see the curves of the maximum deviatoric stress and the maximum pressure in Fig. 5 for examples). With the sustained rise in spin rate, the surface slope at mid-latitudes of the rubble pile exceeds its local friction angle. As a result, some surface particles cannot rest on the surface anymore and landslides occur (Fig. 6(1d)–(4d)). Analogous to the surface avalanche process described in Harris et al. (2009), the mobilized materials flow to the equatorial region, and are shed and leave the asteroid surface. The orbital motion of the shed mass near the equatorial plane results in variations in the axis ratio $b/a$, as shown in the lower-middle frame in Fig. 5.

The failure mode of the HCP configuration is a clear surface-shedding mode. The evolution of surface materials is consistent with the theoretical research carried out by Scheeres (2015), where the asteroid is assumed to be a rigid sphere covered by cohesionless regolith. Given that the motion of internal particles in the HCP structure is restricted by their neighbors while the surface particles have more space to move freely, the HCP structure can serve as a rigid core with a weak surface shell, giving rise to surface failure, as inferred by Hirabayashi et al. (2015). Our SSDEM simulations also confirm that the surface-failure mode of the HCP configuration is its inherent characteristic due to the crystal structure regardless of the interparticle friction (see Section 4.4).

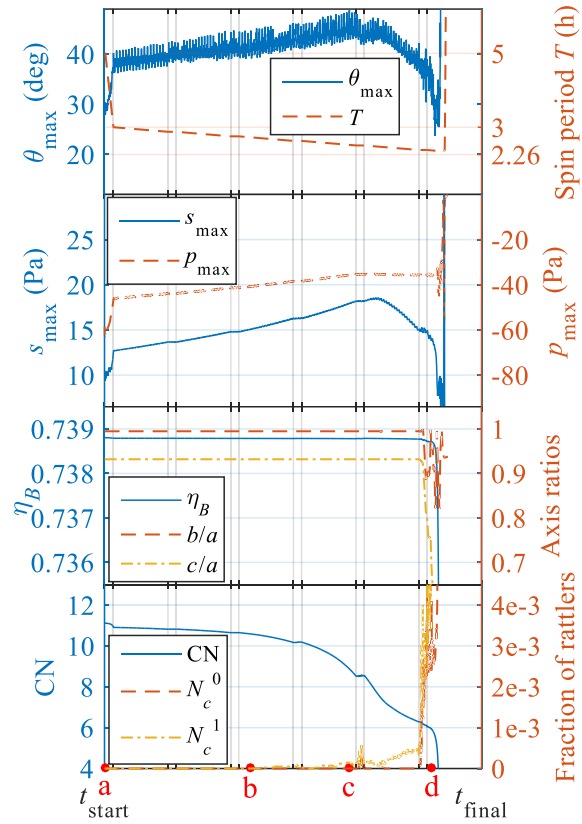

**Figure 5** HCP rubble pile simulation: evolution of the maximum mobilized friction angle and spin period (top frame), maximum deviatoric stress and maximum pressure (upper-middle frame), packing efficiency and axis ratios (lower-middle frame), and coordination number and fraction of rattlers (bottom frame) during the spin-up process. The distribution of the stress-state variables at the times labeled a, b, c, and d on the bottom axis are shown in Fig. 6. The vertical grid is used to indicate the spin-up path, i.e., $t_{inter}$ and the rest-time intervals.

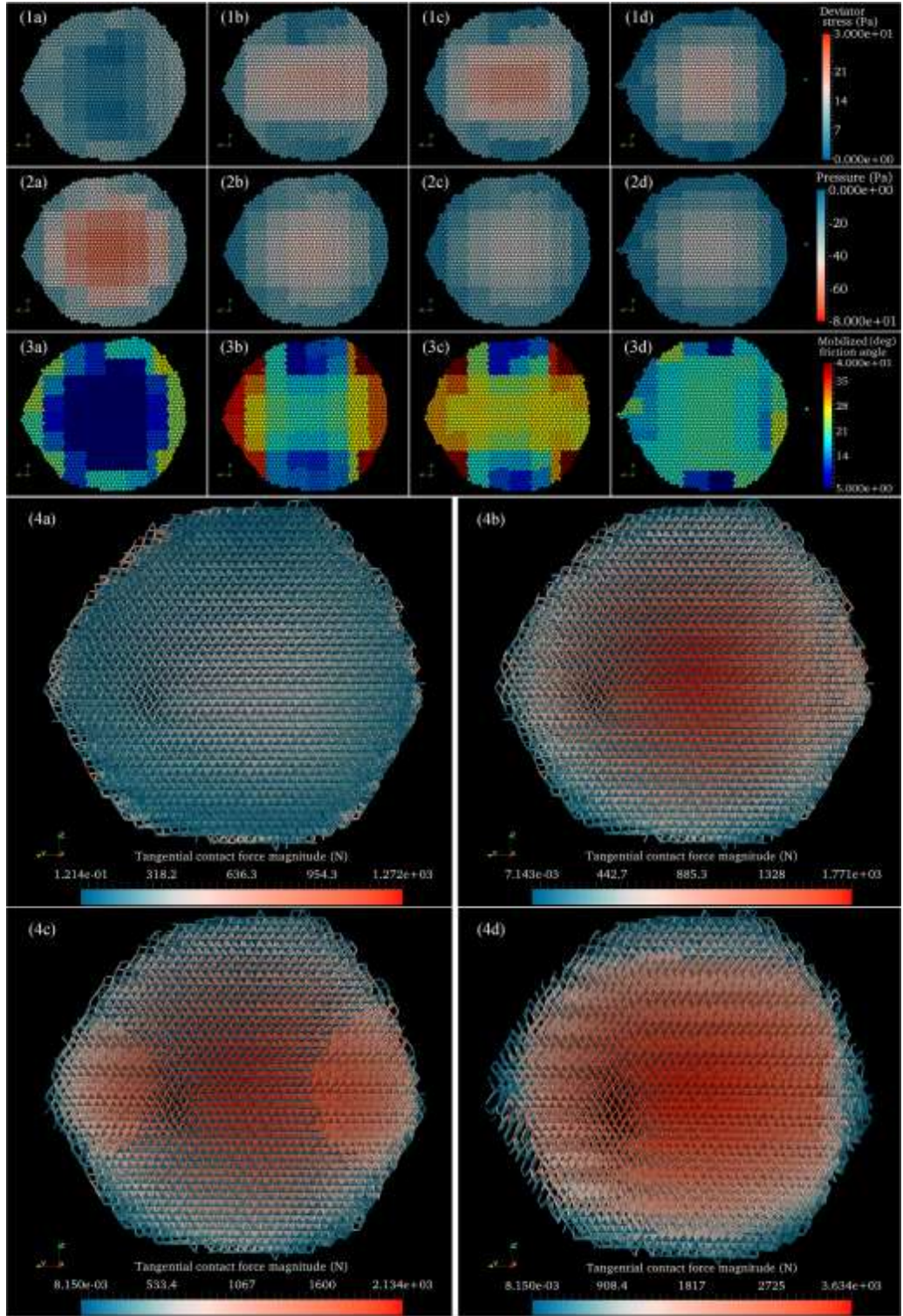

**Figure 6** HCP model simulation: stress-state variables and force networks over a cross-section parallel to the maximum moment of inertia axis at different stages of the spin-up process. Letters a–d correspond to the specific times indicated at the bottom of Fig. 5. The patches evident in the first three rows represent the RVEs in these cross-sections. Note that the color scale for visualizing the force networks corresponds to the magnitude range of tangential contact forces at each specific time.

Owing to the geometrical effects of particle interlocking, the HCP configuration can resist much higher shear stress than a random packing of the same material (Walsh et al., 2012). Using the method introduced in Section 3.3.3, the friction angle for the HCP model is about 43° with a spin limit of $T_{c,3} = 2.40$ h. In fact, due to the geometric interlocking, the actual friction angle for a crystallized structure could be higher (see the curve of $\theta_{max}$ in Fig. 5). The shear strength that the geometric interlocking produces will decrease with the breakage of force chains. As a result, a local region in the HCP structure would fail once its local friction angle has been reduced to below its local mobilized friction angle.

From the lower-middle frame in Fig. 5, the first and second critical periods are 2.38 h and 2.36 h, respectively. By stopping rotational acceleration at various spin rates, the spin period where global failure of the entire body occurs is ~2.37 h. When the body is spun to a spin period between 2.37 h and 2.4 h, only local landslides and surface shedding are observed. The implications of these critical spin limits are discussed in more detail in Section 5.1.

4.1.2 Monodisperse random-close-packing configuration (RCP)

The random close packing leads to a different evolution of stress states compared to the HCP case, as shown in Figs. 7 and 8. From Fig. 8(4a)–(4d), there is a clear change in the orientation of the strong force chains, which are the force chains carrying more than the average load. As would be expected for this random configuration, the distributions of contact force orientations and magnitudes are uniform at a slow spin (Fig. 8(4a)). Since the centrifugal force in the equatorial plane becomes stronger in the spin-up process, the force network needs to reduce the magnitude of its components in the equatorial plane to retain equilibrium. Meanwhile, the contact force chains along the rotation axis must be strengthened to resist the increasing shear stresses in addition to the gravitational pressure. As a result, the orientations of the strong force chains gradually align parallel to the rotation axis. Unlike in the HCP configuration, the breakage of force chains can occur along a range of directions since the particles are distributed in a random way. Therefore, the topology of the force network remains essentially the same throughout the spin-up process, and $\theta_{max}$, $s_{max}$, and $p_{max}$ can keep increasing during the whole spin-up process (Fig. 7).

Figures 8(3c) and (3d) show that $\theta_{max}$ occurs in the interior and the shape becomes more oblate when the body has been spun past its failure limit, without any landslides and surface shedding, implying the internal structure fails first. The distribution of stress-state variables and the failure mode are consistent with the finite-element analyses of a homogenous structure in Hirabayashi and Scheeres (2015).

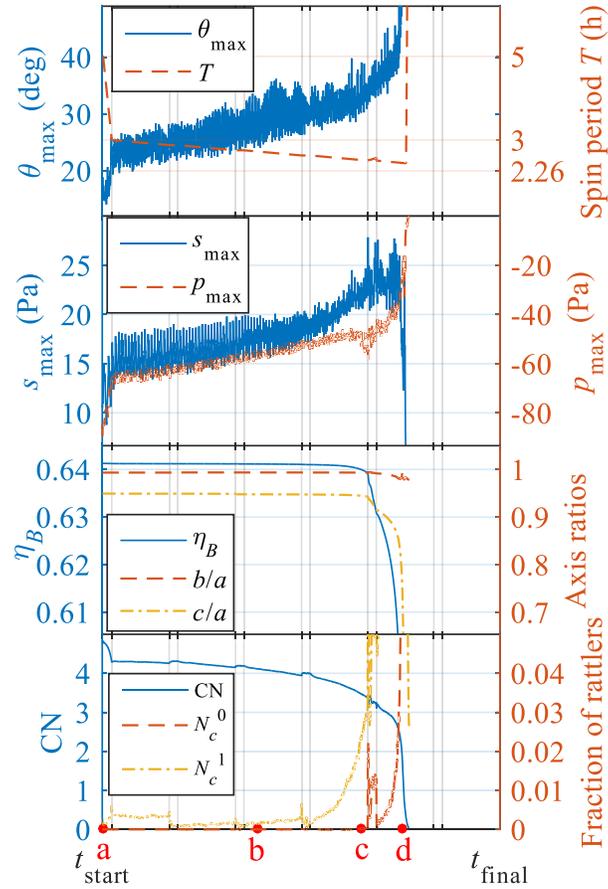

**Figure 7** Same as Fig. 5, but for the RCP model simulation, with the times marked a–d corresponding to the stress state plots in Fig. 8.

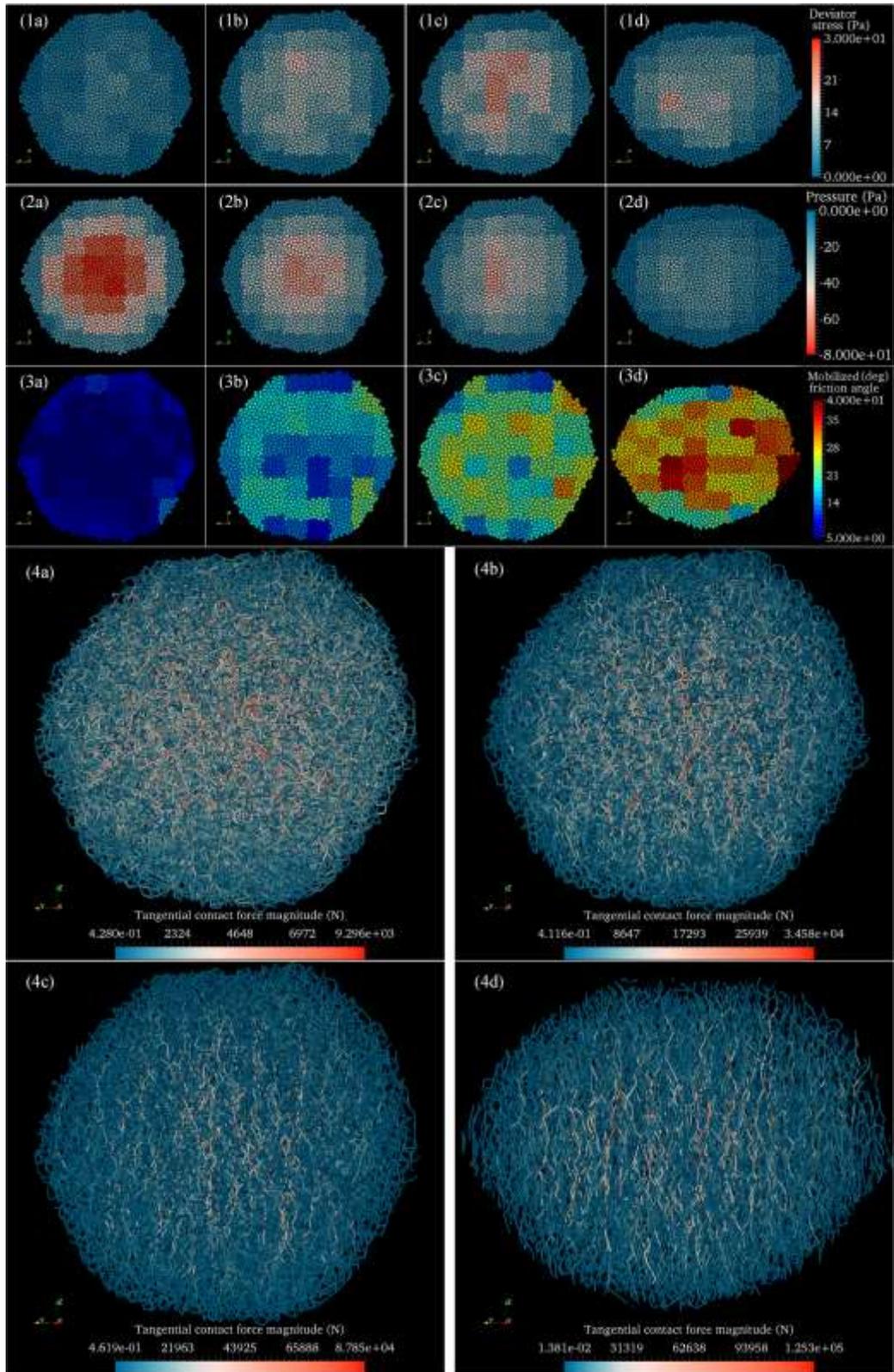

**Figure 8** Same as Fig. 6, but for the RCP simulation, with letters a–d corresponding to the specific times indicated at the bottom of Fig. 7.

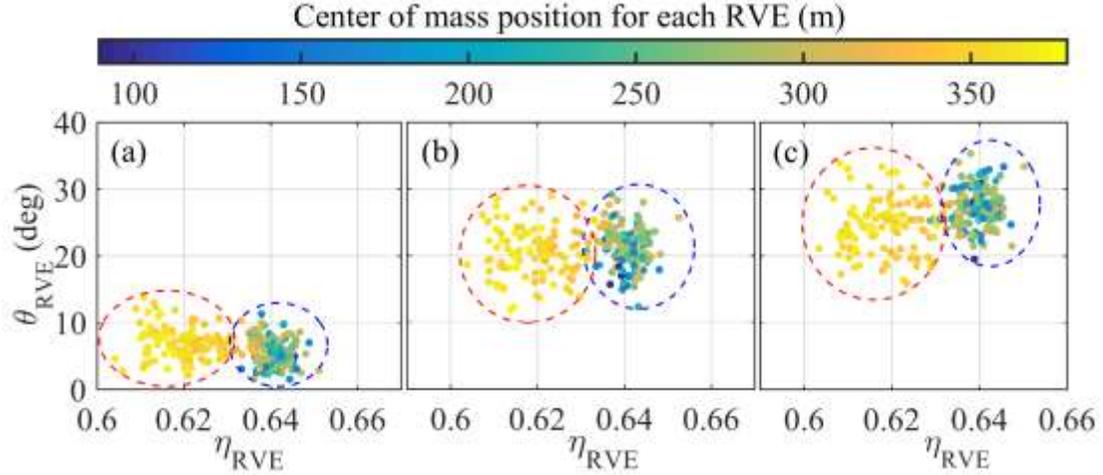

**Figure 9** Mobilized friction angle vs. packing efficiency for each RVE in the RCP case. Color denotes the distance from the aggregate mass center to the center-of-mass position of each RVE. The times corresponding to (a) through (c) are indicated at the bottom of Fig. 7. The red dashed region indicates surface RVEs while the blue dashed region shows interior RVEs.

Figure 9 compares the distribution of the mobilized friction angle for each RVE, $\theta_{RVE}$, to its local packing efficiency, for the RCP model. The packing efficiency for each RVE, $\eta_{RVE}$, can be estimated by averaging the packing fraction of each particle in this RVE (in a way analogous to Eq. (14)). Due to the boundary condition, $\eta_{RVE}$ for surface elements is much lower than that for interior elements, as discussed in Section 3.2. Comparing Figs. 9(a)–(c), we can see a slight relative shift between the mobilized friction angles in the surface RVEs (red circles) and those in the internal RVEs (blue circles). The mobilized friction angles of the internal RVEs increase more rapidly as the angular speed increases. This implies that the internal structure is more sensitive to the increases in the spin state for such a random configuration, which is in agreement with the analytical study on homogeneous, spherical bodies by Hirabayashi (2015). At slow spin rates, the surface region is the most sensitive part to failure in the RCP structure (Fig. 8(3a)), in which case the structure would fail through surface landslides if the material friction angle is low. By monitoring the evolution of the distribution of $\theta_{RVE}$, the RVE where $\theta_{max}$ occurs migrates from the surface to the interior at the spin period of ~3 h, indicating the transition point from the surface failure mode to the internal deformation failure mode. Based on global averaging analyses and finite element modeling, the stress state analyses of Scheeres et al. (2016) on the asteroid Bennu (whose shape is also spheroidal) show a similar transition mechanism between surface failure and interior failure (see their Fig. 22). The consistency reveals that the RCP model may be an appropriate discrete representation of the continuum approach. The reason for this is discussed in more detail in Section 5.1.1. Furthermore, the maximum mobilized friction angle in the internal structure usually occurs at the element whose packing efficiency is relatively low (Fig. 9). This may mean that the rubble-pile structure tends to fail in a low-porosity region.

By stopping the rotational acceleration at various spin rates, we find the third critical spin period for the RCP model is about 2.83 h, below which global structural failure would occur. Using the method introduced in Section 3.3.3, the friction angle of the RCP model is about 30°. When the structure yields, the poles of the rubble pile push inwards toward the center, leading to a more oblate shape. Since no surface shedding occurs in the RCP model, the reshaping behavior is a result of the internal deformation and volumetric dilation process. From the lower-middle frame in Fig. 7, it is evident that the packing efficiency begins to decrease before reshaping occurs, corresponding to first and second critical periods of 2.51 h and 2.53 h, respectively.

4.1.3 Mixed core/shell packing configuration (RCPC)

Even though the previous two models both consist of equal-size particles, the differences in the spatial arrangement of the granular systems lead to two very different failure modes. Given the extra shear strength that arises from geometric interlocking in the HCP structure, we can construct a body with a stronger internal core by placing these core particles in an HCP arrangement, which is what we did for the RCPC case considered here. The evolution of the RCPC configuration progresses in a manner similar to the combination of HCP and RCP structures, as shown in Figs. 10 and 11. First, $s_{max}$ and $p_{max}$ evolve generally in the manner of the HCP configuration. Since the maximum deviatoric stress and the maximum pressure always occur in the internal HCP core (as presented in the first two rows of Fig. 11), $s_{max}$ and $p_{max}$ stop increasing when the corresponding local force network begins to change its topology. $\theta_{max}$ evolves in a more complex way. At low spin rates, the internal core is subject to the highest mobilized friction angle due to the higher internal shear stress, in which case $\theta_{max}$ evolves in the manner of the HCP struc-ture. For the same reason as $s_{max}$ and $p_{max}$, the mobilized friction angle in the internal core will also decrease when the interior begins to change its topology. Once the maximum mobilized friction angle of the internal core is smaller than that of the external shell, $\theta_{max}$ no longer occurs in the interior and instead occurs in the external shell. Thus, $\theta_{max}$ can continually increase in the same manner as the RCP structure after that point. Since the internal region is more sensitive to changes in the spin state for a random packing (as discussed in Section 4.1.2), the position of the RVE where $\theta_{max}$ is obtained gradually moves to the border of the HCP core and the RCP shell. The evolution of the $\theta_{max}$ position is seen in the third row of Fig. 11. Furthermore, the characteristics of both HCP and RCP configurations are also observed in the force network of the RCPC model, such as the breakage of horizontal force chains and the orientations of the strong force chains.

Surface shedding and internal deformation are both observed for this RCPC model. From Figs. 11 (1d)–(4d), we can see some orbiting material near the equator. However, the packing efficiency of the internal HCP core remains constant for a while after the external shell fails. This implies that the flow of particles is restricted to the external shell, and the HCP core can serve as a rigid core before global disruption occurs. The existence of a strong core frustrates the equatorial elongation, resulting in surface flow and mass shedding. This failure mode with a strong core is consistent with the analyses of Hirabayashi et al. (2015). Furthermore, our study reveals that the part subject to the highest shear stress in this composite structure changes with the spin rate, and finally appears at the border of the two configurations, leading to a strong shear effect between the internal core and the external shell. Therefore, in addition to the surface shedding behavior, the granular assembly of the external shell deforms and redistributes itself above the surface of the border.

The third critical spin period for the RCPC model is about 2.76 h with a friction angle of about 32°, which is right in between that of the HCP model and that of the RCP model. Comparison between the distributions of deviatoric stresses and mobilized friction angles of the RCP model (Fig. 8) and those of the RCPC model (Fig. 11) show that the HCP core not only enhances the shear resistance of the central region but also helps to ease the shear stresses in the external shell. Due to the fact that the friction angle of the HCP core is much higher than that of the RCP shell, the external shell fails first when the shear stresses exceed its own failure limit (i.e., 30° for the RCP structure), even though $\theta_{max}$ (which is about 32° at the failure point) of the whole structure still occurs at the internal core. Therefore, the RCPC configuration can be stable at a higher spin rate compared to the pure random packing case.

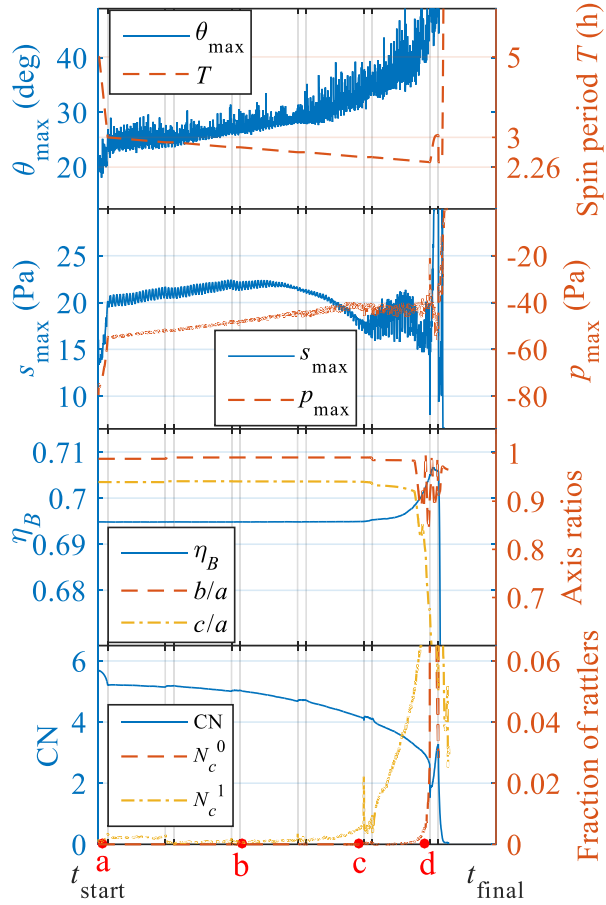

**Figure 10** Same as Figs. 5 and 7, but for the RCPC simulation, with the times marked a–d corresponding to the stress state plots in Fig. 11.

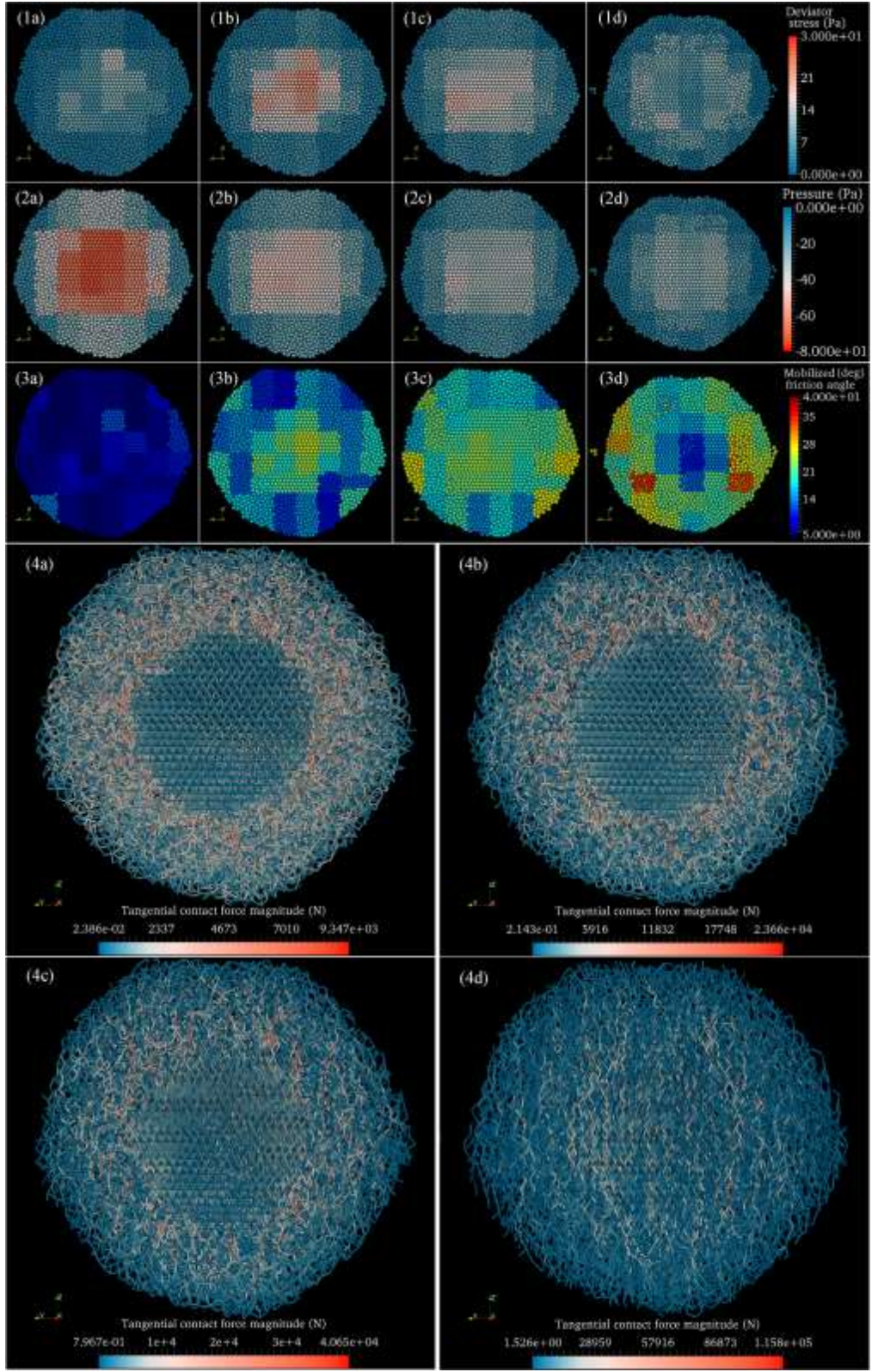

**Figure 11** Same as Figs. 6 and 8, but for the RCPC simulation, with letters a–d corresponding to the specific times indicated at the bottom of Fig. 10.

Unlike any of the previous models, the first and second critical periods for the RCPC model have the same value, of 2.46 h. After the external shell yields, some surface materials are shed and leave the main body, leading to an increase in the volume fraction of the HCP core. Since the local filling fraction of the HCP configuration (~0.74) is much higher than that of the random packing (~0.64), the packing efficiency of the main body increases with the external shell shedding, as shown in the lower-middle frame of Fig. 10. In this case, the variations in the packing efficiency may not appropriately represent the internal deformation. By tracking the local packing fraction of the external shell, we observed that the deformation and the surface shedding almost occur at the same time, and are both triggered by the redistribution and flow of particles in this shell.

4.1.4 Polydisperse packing configuration (PP1)

The previous models were constructed using monodisperse particles, which is a very idealized situation. This and the following cases use polydisperse aggregates. We constructed our first polydisperse packing model (PP1) using particles whose radii follow a power law with an exponent of –3 and lower and upper cutoff values of 4 m and 32 m, respectively. The evolution and distribution of the stress states behave in a similar manner to the RCP configuration, as shown in Figs. 12 and 13. The orientations of the strong force chains also gradually align parallel to the rotation axis during the spin-up process. However, compared with the RCP model, the fraction of the strong force chains is smaller and the whole network appears to be highly heterogeneous in the PP1 model. An explanation for this is that the force chains in a highly polydisperse granular assembly are mainly maintained by large particles. Figure 14(A) shows the relation between the pressure magnitude and the particle radius at time b (see Fig. 12). The pressure $p_p$ is calculated based on the stress tensor of a single particle using Eq. (15). The majority of particles (especially the small ones) in this polydisperse model carry less than the average load. Meanwhile, there are a significant number of small particles subject to very small pressure in the interior region; these may be rattlers with only one contact (see the blue dots in the lower left part of Fig. 14(A); compare the bottom frame in Fig. 12 with that in Fig. 7—the fraction of rattlers for the PP1 model is much higher). This implies that there are numerous void spaces permitting the migration of small particles in the interior region. Although the range of movement is very limited, the variations in the contact forces that result from the oscillations of these smaller particles are considerable. As shown in Fig. 14(A), the magnitude of pressure for the small particles can vary from $10^{-2}$ Pa to $10^4$ Pa. For the same reason, the curves of $p_{max}$ and $s_{max}$ show significant oscillations (see the upper-middle frame in Fig. 12). Conversely, due to geometric interlocking between big particles, the relative movements between big particles are much tinier. The range of pressure magnitude shrinks for larger particles, and the magnitude of pressure becomes inversely proportional to the particle's distance from the aggregate center (Fig. 14(A)).

Figures 13(3c) and (3d) show that the maximum mobilized friction angle occurs in the interior and the shape becomes more oblate when the body has been spun past its failure limit, implying the internal structure fails first. However, differing from the RCP model, surface shedding is observed in this case. The orbital motion of the shed mass near the equatorial plane results in variations in the axis ratio $b/a$, as shown in the lower-middle frame in Fig. 12. From animations of simulation behavior, we find that the landslide starts with some small surface particles flowing to the equatorial region and leaving the surface. Following this, some nearby big surface particles are able to move and be shed. That is to say, the size heterogeneity makes room for surface material movement, while the geometric interlocking caused by the monodispersity restricts the motion of the surface particles in the RCP model. Another possible reason to explain the surface shedding behavior is the difference in shear strength between the surface and interior. The high internal packing efficiency (~0.82) enhances the shear resistance of the interior, which may slow down the deformation process. And the surface shear strength (reflected in the value of the surface angle of friction), which depends on the material parameters and geometric effects, would be less than the internal shear strength. In this case, the surface material would fail when the local mobilized friction angle exceeds its surface angle of friction. Therefore, even though the interior is subject to the highest shear stresses, the difference in the surface and interior shear strengths allows surface shedding to occur without significant internal deformation.

Using the method introduced in Section 3.3.3, the friction angle for the PP1 model is ~39° and the spin limit against global structural failure $T_{c,3}$ = 2.53 h. Compared to the monodisperse random packing model (RCP), the size heterogeneity and the higher packing efficiency of the PP1 model indeed improve the shear strength of the rubble-pile structure. Note that the packing efficiency is a function of the size heterogeneity in this study since the specimen is generated through a shaking process; that is, higher size heterogeneity leads to higher packing efficiency. In fact, asteroids in space are naturally subject to various vibration excitations (e.g., seismic shaking caused by meteorite impacts or planetary encounters) and size segregation within the granular materials may arise from these processes (Asphaug et al., 2001). Vibration-induced size sorting can result in a more compacted structure. Therefore, a highly polydisperse aggregate with a high packing fraction can serve as an appropriate representation of a rubble-pile asteroid. Once the actual porosity of a rubble-pile asteroid is obtained, some characteristics of its particle size distribution could be inferred. From the lower-middle frame in Fig. 12, consistent with the internal failure mode, the interior packing efficiency begins to decrease before reshaping occurs, with corresponding first and second critical periods of 2.41 h and 2.44 h, respectively.

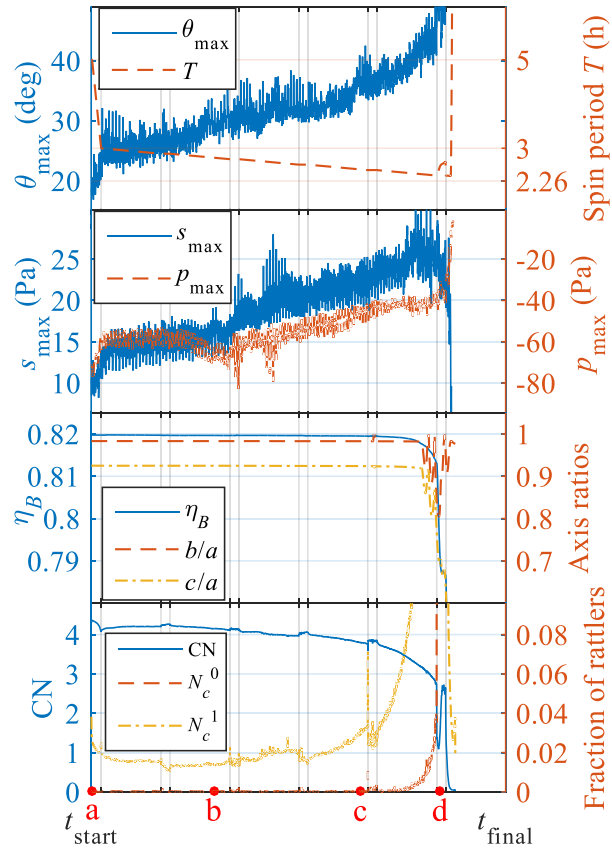

**Figure 12** Same as Figs. 5, 7 and 10, but for the PP1 simulation, with the times marked a–d corresponding to the stress state plots in Fig. 13.

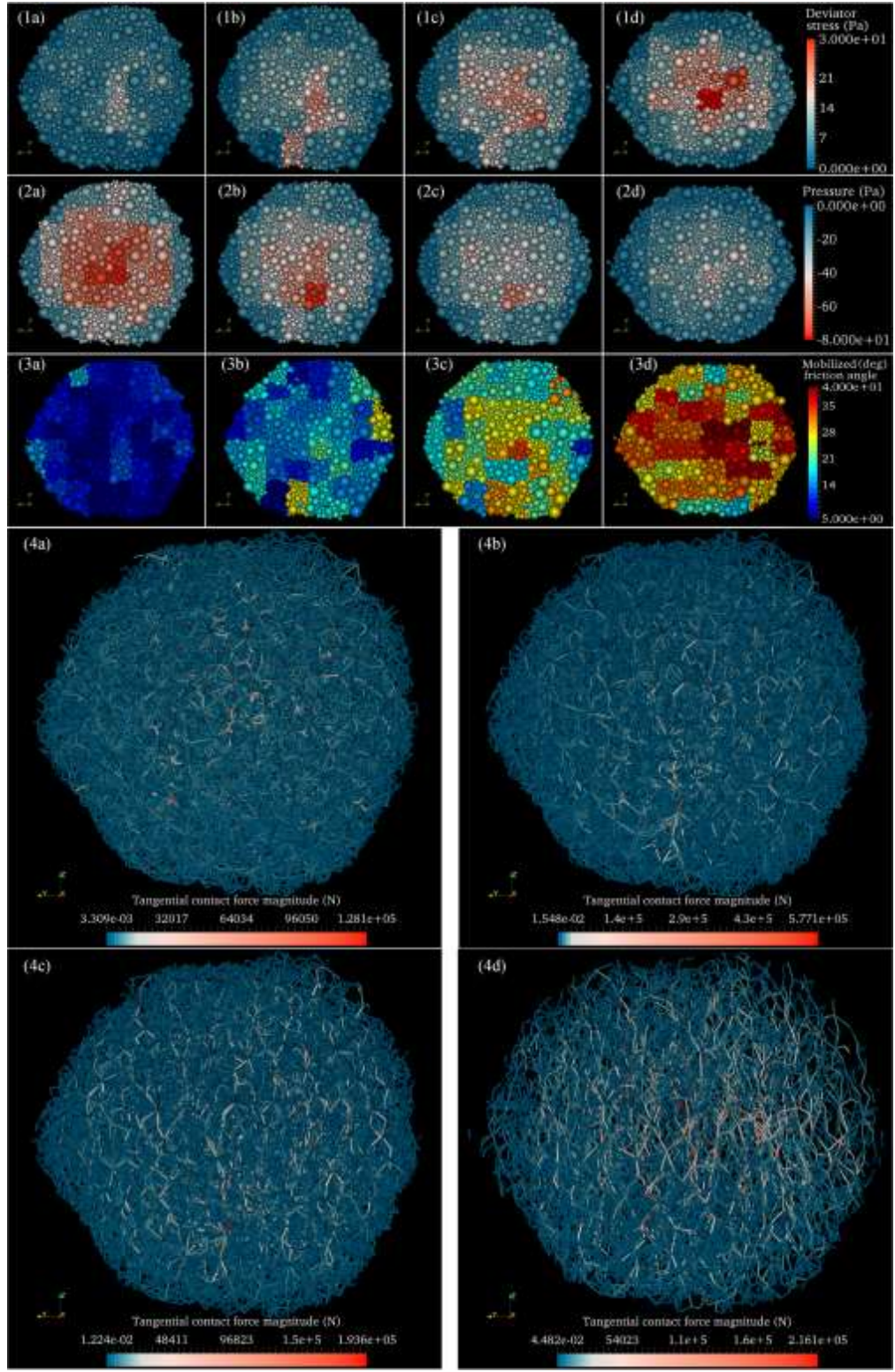

**Figure 13** Same as Figs. 6, 8, and 11, but for the PP1 simulation, with letters a–d corresponding to the specific times indicated at the bottom of Fig. 12.

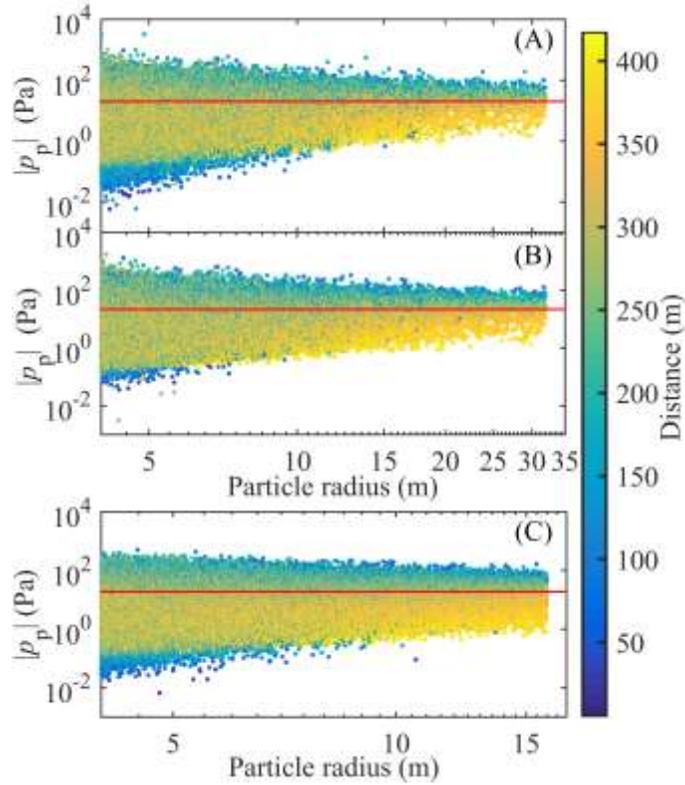

**Figure 14** Distribution of pressure magnitude acting on each particle as a function of the particle radius at time b (see Figs. 12, 15, and 17): (A) PP1 model; (B) PPC model; (C) PP2 model. The red line indicates the magnitude of the average pressure for each case. Color denotes the distance from the aggregate mass center to the particle center.

4.1.5 Polydisperse packing configuration (PP2)

The size distribution of the second polydisperse packing model (PP2) follows a power law with an exponent of –3 with lower and upper cutoff radii of 4 m and 16 m. The evolution of this polydisperse packing model progresses in a manner similar to that of PP1 (Figs. 15 and 16). Given that its radius ratio (i.e., the ratio of the maximum particle radius to the minimum particle radius in the assembly) is smaller than in the previous model, the distribution of the strong force chains is less heterogeneous in the PP2 model. The amplitude of the stress states curves (see the upper-middle frame in Fig. 15) and the span of the particles' pressure (see Fig. 14 (C)) become narrower as well.

As for the PP1 model, the failure of the PP2 structure is triggered by internal structural deformation. As shown in the lower-middle frame of Fig. 15, the internal packing efficiency begins to decrease before reshaping occurs, corresponding to first and second crit-ical periods of 2.41 h and 2.44 h, respectively. The angle of friction for the PP2 model is about 38° with a third critical spin period of $T_{c,3}$ = 2.54 h. These results imply that the shear strength of the PP2 model is slightly lower than that of the PP1 model, supporting the idea that the increased size heterogeneity associated with a higher packing efficiency leads to higher shear strength. In the current study, we only investigate the size distribution with one plausible power law. A systematic study on the effect of size distribution will be conducted in a further study.

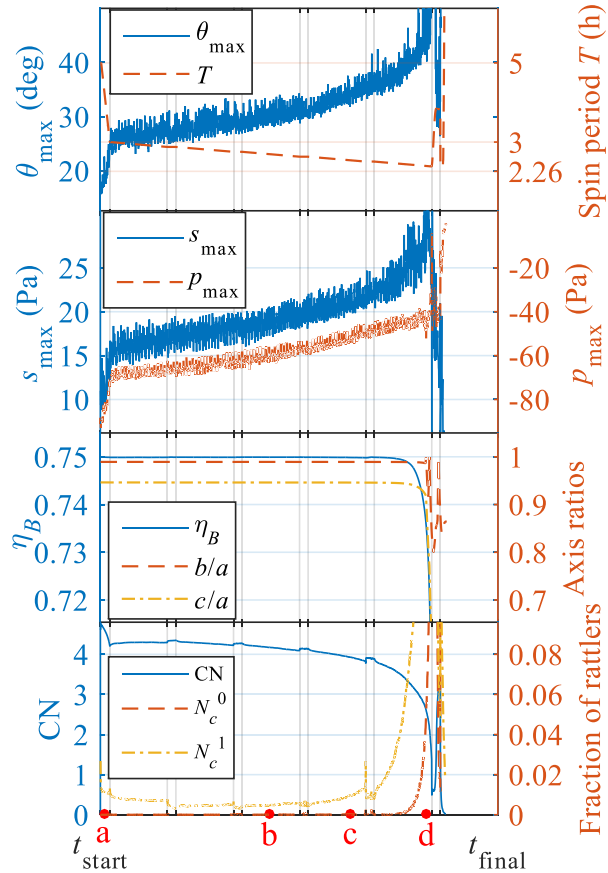

**Figure 15** Same as previous evolution figures, but for the PP2 simulation, with the times marked a–d corresponding to the stress state plots in Fig. 16.

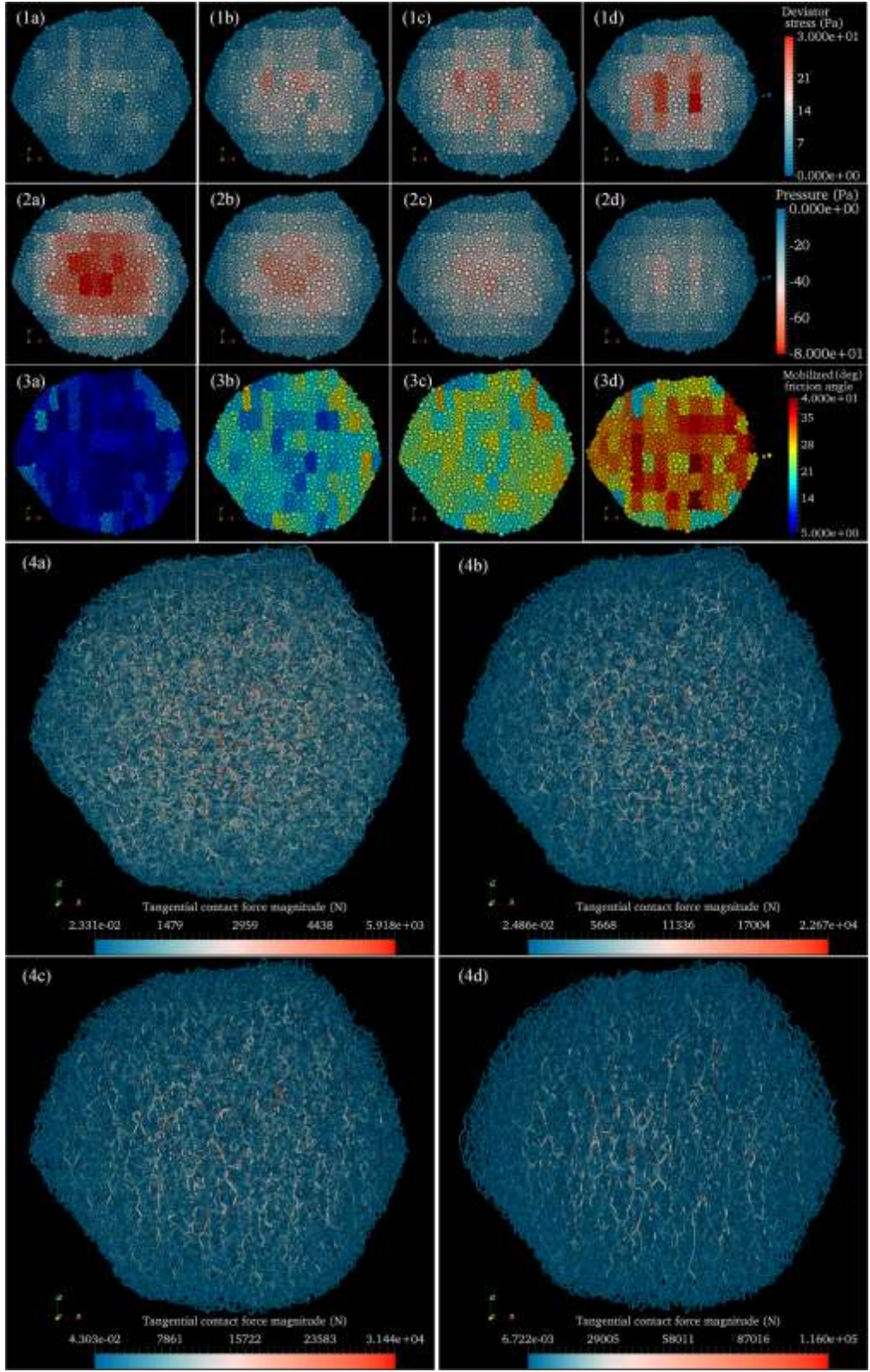

**Figure 16** Same as previous cross-section figures, but for the PP2 simulation, with letters a–d corresponding to the specific times indicated at the bottom of Fig. 15.

### 4.1.6 Polydisperse dense-core configuration (PPC)

The previous mono-density structures may not be suitable for the majority of asteroids. As a final study case, we constructed a bi-density model with a 200-m-radius denser core using the original PP1 configuration to investigate the effect of density heterogeneity. The fraction of rattlers in the denser core is significantly less than that of the PP1 model, resulting in a more uniform pressure distribution within it (Fig. 14 (B)). The evolutions of the stress variables in the PPC model are similar to those of the PP1 model (Fig. 17), while the distributions are quite different (Fig. 18). During the spin-up process, shear stresses acting on particles increase as the centrifugal forces grow. By imposing a higher gravitational pull among these particles, the denser core helps to ease the shear stresses in the interior. Therefore, as shown in Fig. 18, $\theta_{max}$ always occurs in the external shell, which will fail first when $\theta_{max}$ exceeds the friction angle.

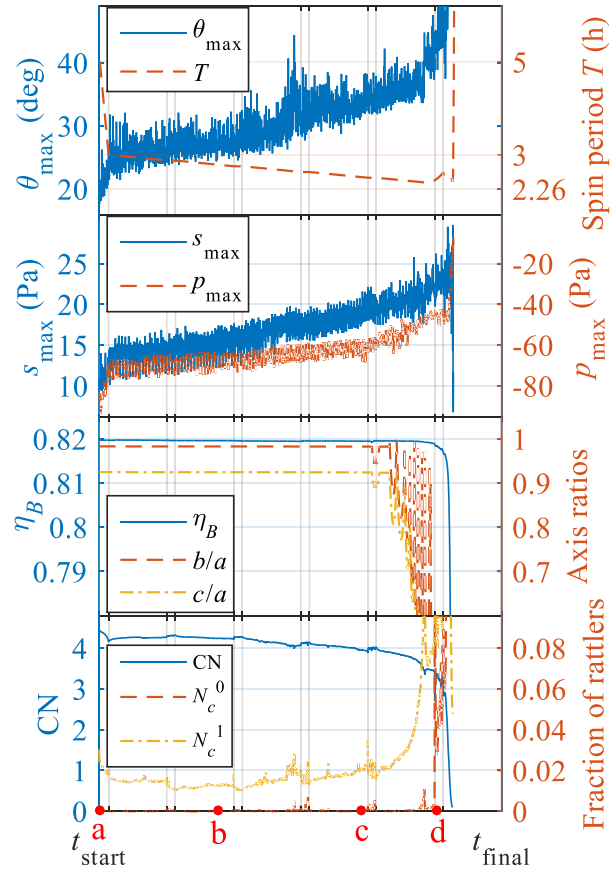

**Figure 17** Same as previous evolution figures, but for the PPC simulation, with the times marked a–d corresponding to the stress state plots in Fig. 18.

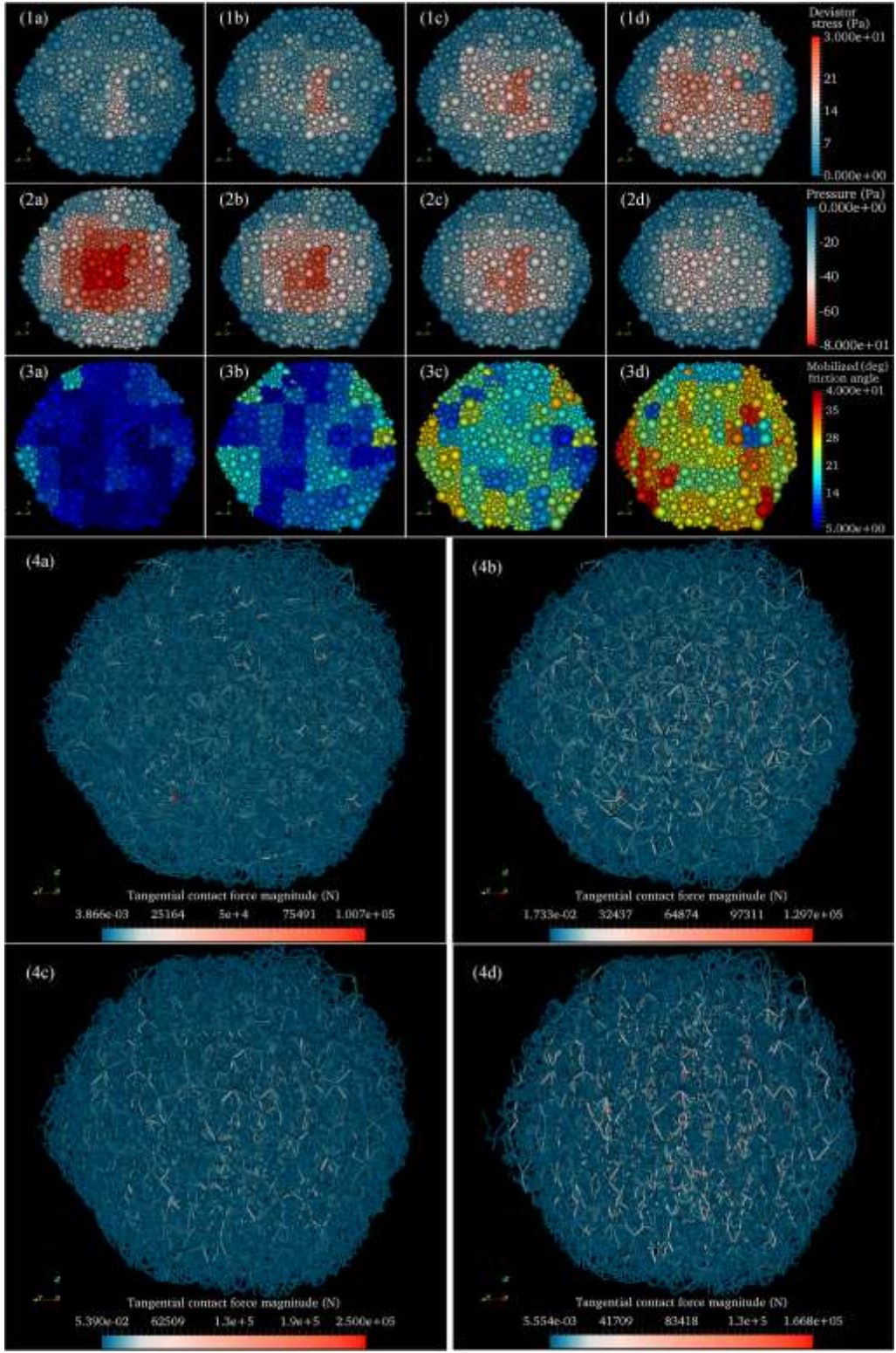

**Figure 18** Same as previous cross-section figures, but for the PPC simulation, with letters a–d corresponding to the specific times indicated at the bottom of Fig. 17.

The friction angle for this model is about 39°, the same as the PP1 model, but with a smaller third critical spin period of $T_{c,3}$ = 2.49 h. Since the shear stress in the external shell grows more slowly with spin rate than that in the interior, there is a delay in the onset of failure of the external shell compared to the PP1 model. For this reason, the PPC model can be stable at a higher spin rate.

As expected from the distribution of the mobilized friction angle, the failure mode of the PPC configuration is a clear surface-shedding mode. Different from the uniform density models, the axis ratios begin to vary before internal deformation occurs, as shown in the lower-middle frame in Fig. 17, corresponding to first and second critical periods of 2.46 h and 2.41 h, respectively. The variations in the axis ratio $b/a$ reveal the orbital motion of the shedding mass near the equatorial plane. Since the gravitational pulls and contact forces between particles in the low-density shell become smaller, the PPC model starts surface shedding at a higher spin period than the PP1 model.

4.2 Effect of bulk density

Obviously, none of the six models can be stable at the current observed spin period of 2.26 h of the Didymos primary with the nominal mass, $5.23 \times 10^{11}$ kg. In fact, the estimates of the mass and bulk density of the primary are subject to a considerable error, as indicated in Table 1. The actual bulk density may be as high as 2730 kg/m$^3$ (i.e., one standard deviation above the nominal density). The dependence of the critical spin rate $\Omega_c$ on the bulk density $\rho_B$ of a rubble pile is given as

$$\Omega_c = \kappa \sqrt{4\pi G \rho_B / 3}, \tag{22}$$

where $G$ is the universal gravitational constant and $\kappa$ is a constant that depends on model and material properties. A higher $\kappa$ indicates a stronger structure. This dependence has been verified by theoretical analyses (e.g., Holsapple, 2001; Sharma et al., 2009) and is consistent with DEM simulations (e.g., Sánchez and Scheeres, 2012). Equation (22) shows that one way to improve the spin limit of a rubble pile is to increase its bulk density.

In this study, we define the *critical bulk density*, $\rho_B^{c,i}$ (where $i$ = 1, 2, and 3), as the bulk density for which the corresponding critical spin period of the rubble pile is 2.26 h. In order to test the critical bulk density for all six models, multiple simulations were run with various bulk densities using the nominal spin-up path and material parameters. To be consistent with the bulk density of the Didymos primary in the nominal case, the bulk density $\rho_B$ is calculated as $2100(M_a/M_P)$ kg/m$^3$, where $M_a$ is the total mass of the aggregate and $M_P$ is the nominal mass of the primary (i.e., $5.23 \times 10^{11}$ kg). Rubble-pile models with different bulk densities are achieved by changing $M_a$, without changing the shape. In reality, the main uncertainty in the bulk density estimate is the error on the shape, not the mass, which is better constrained. However, given that the shear strength of a cohesionless body is independent of its size (Holsapple, 2001; Hirabayashi, 2015), we decide to fix a reference model based on shape, and vary the mass, which is easier than constructing various shape models and can show the same effect of the bulk density.

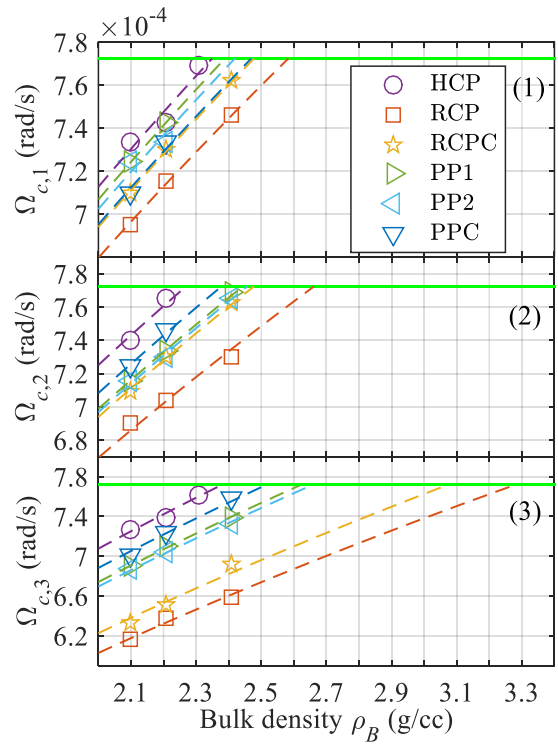

**Figure 19** Critical spin rates as a function of bulk density for the six rubble-pile models: (1) the first critical spin rate (deduced from variations in axis ratios); (2) the second critical spin rate (deduced from variations in packing efficiency); and (3) the third critical spin rate (deduced from the Mohr-Coulomb yield criterion). The dashed lines are fits to Eq. (22) for the corresponding markers with the same color. The horizontal green line marks the current observed spin rate of the Didymos primary.

Figure 19 shows the relation between the critical spin rates and the bulk densities for the various rubble-pile configurations we tested. The three definitions of critical spin rates all show a square-root dependency on the bulk density, but with different prefactors $\kappa_i$ (where $i$ = 1, 2, and 3). The critical bulk density can be estimated through fitting the data according to Eq. (22). Table 4 summarizes the critical bulk densities and the proportionality constants. As expected, the polydisperse random packing model can be stable at a lower bulk density than the monodisperse random packing model, and the HCP model is capable of maintaining overall creep stability at an even lower bulk density. We discuss the implications of these results in Section 5.2.

**Table 4** Critical bulk densities of rubble-pile models. See main text (Section 4.2) for definitions of the symbols.

| Rubble-pile Model | $\rho_B^{c,1}$ (kg/m$^3$) | $\kappa_1$ | $\rho_B^{c,2}$ (kg/m$^3$) | $\kappa_2$ | $\rho_B^{c,3}$ (kg/m$^3$) | $\kappa_3$ |
|---|---|---|---|---|---|---|
| HCP | 2349 | 0.953 | 2269 | 0.970 | 2383 | 0.946 |
| RCP | 2583 | 0.909 | 2662 | 0.895 | 3284 | 0.806 |
| RCPC | 2478 | 0.928 | 2478 | 0.928 | 3077 | 0.833 |
| PP1 | 2388 | 0.945 | 2443 | 0.935 | 2626 | 0.902 |
| PP2 | 2419 | 0.939 | 2457 | 0.932 | 2662 | 0.895 |
| PPC | 2469 | 0.930 | 2378 | 0.947 | 2519 | 0.920 |

4.3 Effect of spin-up path

It is well known that a difference in loading rate leads to a different speed of stress-wave propagation in granular materials (Russell et al., 2014). Furthermore, the tangential force and the rotational torques in our contact model could be affected by the relative rotational and translational velocities of the particles as well as the history of their contacts. Our previous study has shown strain-rate (loading rate) strength effects and crack-propagation effects in Brazilian disk tests (Schwartz et al., 2013). Therefore, a substantial change in the creep stability of a spinning rubble pile due to variations in the loading rate might be observed. Although previous numerical studies on spin-up of rubble piles (e.g., Sánchez and Scheeres, 2016) generally ignore the effect of spin-up path by assuming quasi-static loading, it is necessary to investigate this effect in the current study with the aim of clarifying possible physical properties for the Didymos primary, for which the timescale of the loading process exceeds $10^4$ years (Rubincam, 2000). Since the time step for an SSDEM simulation is usually on the order of microseconds or even less, existing processing power cannot support a computation on such a scale even with parallel techniques.

Therefore, exploring the effect of spin-up path could give the dependence of the spin limits on the timescale of loading, and shed light on the actual behavior of the asteroid over time.

The failure mode for the six models is one of surface shedding, internal deformation, or a combination of both. Since the HCP configuration and the PP2 configuration exhibit the features of all failure modes, we take these two models as representative cases for this topic. Table 5 summarizes the parameters of spin-up paths investigated in this study, for which path No. 1 is the nominal spin-up path.

**Table 5** Summary of spin-up paths. See Section 2.3.4 for definitions of the spin-up path parameters.

| Path No. | $T_{start}$ (h) | $T_{inter}$ (h) | $T_{final}$ (h) | $t_{start}$ (s) | $t_{inter}$ (s) | $t_{final}$ (s) | Settle-down times | $\Delta t_{rest}$ (s) |
|---|---|---|---|---|---|---|---|---|
| 1 | 5.0 | 3.0 | 2.26 | 700 | 7000 | 320000 | 5 | 7000 |
| 2 | 5.0 | 3.0 | 2.26 | 700 | 7000 | 32000 | 0 | 0 |
| 3 | 5.0 | 4.0 | 2.26 | 700 | 7000 | 32000 | 0 | 0 |

As shown in Figs. 20 and 21, the evolution of the maximum mobilized friction angle as a function of the spin period follows the same trend and almost the same value before the body fails, regardless of the spin-up path. The third critical spin period deduced from the Mohr-Coulomb yield criterion, $T_{c,3}$, is the same for the different spin-up paths, as shown in Table 6. These results hold true for a range of bulk densities investigated in this study. This implies that the stress states in a geo-statically stable rubble-pile structure mainly depend on its current spin state and not on how fast the spin is increased within the range of loading rates achieved in this study. [12]

**Table 6** Critical spin periods under different spin-up paths for the HCP and PP2 models. Here $\rho_B$ is the bulk density of the rubble pile. See Section 3 for the definitions of the critical spin periods, $T_{c,1}$, $T_{c,2}$ and $T_{c,3}$.

| Rubble-pile Model | $\rho_B$ (kg/m³) | Path No. 1 | | | Path No. 2 | | | Path No. 3 | | |
|---|---|---|---|---|---|---|---|---|---|---|
| | | $T_{c,1}$ (h) | $T_{c,2}$ (h) | $T_{c,3}$ (h) | $T_{c,1}$ (h) | $T_{c,2}$ (h) | $T_{c,3}$ (h) | $T_{c,1}$ (h) | $T_{c,2}$ (h) | $T_{c,3}$ (h) |
| HCP | 2100 | 2.38 | 2.36 | 2.40 | 2.30 | 2.27 | 2.40 | 2.26 | 2.26 | 2.40 |

---

[12] Note that this study only investigated monotonic spin-up paths. The same conclusion may not apply for non-monotonic spin-up paths caused by the stochastic YORP effect, which may contain several spin-up and spin-down stages (Statler, 2009). History dependence may be found when the loading paths are non-monotonic, which is left for future study.

| | | | | | | | | | |
|---|---|---|---|---|---|---|---|---|---|
| | 2208 | 2.35 | 2.28 | 2.36 | 2.26 | 2.26 | 2.36 | 2.26 | 2.26 | 2.36 |
| | 2309 | 2.27 | 2.26 | 2.29 | 2.26 | 2.26 | 2.29 | 2.26 | 2.26 | 2.29 |
| | 2100 | 2.41 | 2.44 | 2.54 | 2.33 | 2.37 | 2.54 | 2.30 | 2.35 | 2.54 |
| PP2 | 2208 | 2.38 | 2.39 | 2.48 | 2.28 | 2.31 | 2.48 | 2.26 | 2.29 | 2.48 |
| | 2409 | 2.26 | 2.28 | 2.38 | 2.26 | 2.26 | 2.38 | 2.26 | 2.26 | 2.38 |

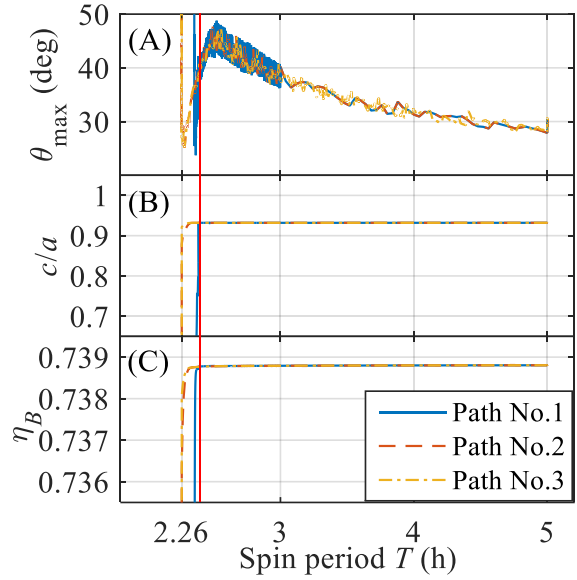

**Figure 20** Comparison of simulations with different spin-up paths for the HCP model with $\rho_B = 2100$ kg/m$^3$ as a function of spin period: (A) maximum mobilized friction angle; (B) axis ratio $c/a$; and (C) packing efficiency. The vertical red line indicates the value of $T_{c,3}$ for this model, which is essentially the same for all 3 spin-up paths.

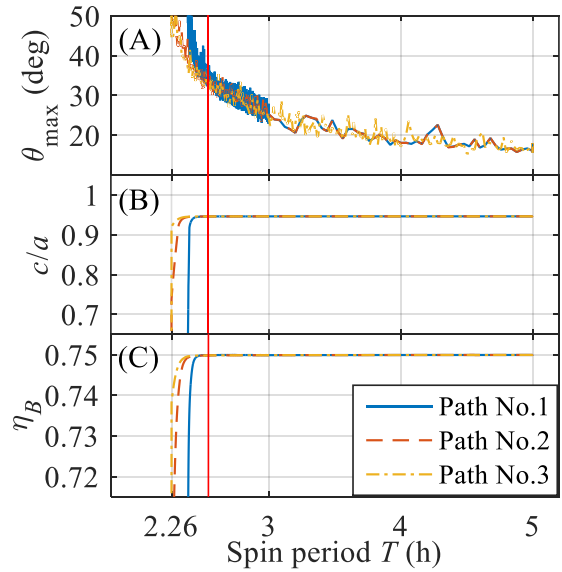

**Figure 21** Same as Fig. 20, but for the PP2 model.

However, the spin rate that triggers the onset of reshaping and surface shedding or internal deformation depends on the spin-up path: the faster the spin rate is increased, the lower the corresponding critical spin period (i.e., $T_{c,1}$ or $T_{c,2}$). This phenomenon is related to the delayed mechanical response of granular materials, which has been reported in the literature (e.g., Oda, 1972; di Prisco and Imposimato, 1996; Ghiabi and Selvadurai, 2009). Naturally, the mechanical response of such materials cannot be immediate. After inducing finite load increments (i.e., incremental centrifugal forces in this study) within a granular assembly, the stress distribution at the microscale changes with time and the microstructural arrangement continuously adjusts. This progressive reconstruction with increasing stresses is mainly caused by sliding between particles in contact and partly caused by the rotation of particles, which are well modeled within our SSDEM implementation. When the shear stress is below the material shear resistance, the vibrations of particles due to the reconstruction process are quickly damped out, and the whole granular system reaches a new equilibrated distribution to balance the external applied macro-load increments; otherwise, creep deformation gradually develops, leading to a continual increase in the shear stress with time, until the whole structure eventually disrupts.

Creep motion has been observed in most sheared granular systems, and occurs at time scales of hours to days for granular materials in laboratory investigations (Ghiabi and Selvadurai, 2009). Figure 22 compares the evolution of state variables for the PP2 model with different bulk densities. The cases of $\rho_B = 2409$ kg/m$^3$ and $\rho_B = 2610$ kg/m$^3$ exhibit a clear creep behavior, where the rubble pile can still maintain its shape without exhibiting apparent deformation or surface shedding for a while after it has been spun past its failure limit (recall that the friction angle of the PP2 model is about 38°). Assuming the timescale of the creep behavior is constant for a certain structure during the spin-up processes, the macro-reconfiguration of the structure will occur at a higher spin rate when a shorter spin-up time is applied, as shown in Table 6.

In addition, because of the creep behavior, some simulated bodies that can maintain their shapes at the end of the spin-up process will eventually break up some time after achieving the final spin period, $T_{\text{final}}$. Since the body is forced to stay at $T_{\text{final}}$ after the spin-up process, the first and second critical spin period where the body begins to re-shape/deform are equal to $T_{\text{final}}$, as seen in Table 6. In these cases, the values of $T_{c,1}$ and $T_{c,2}$ are meaningless.

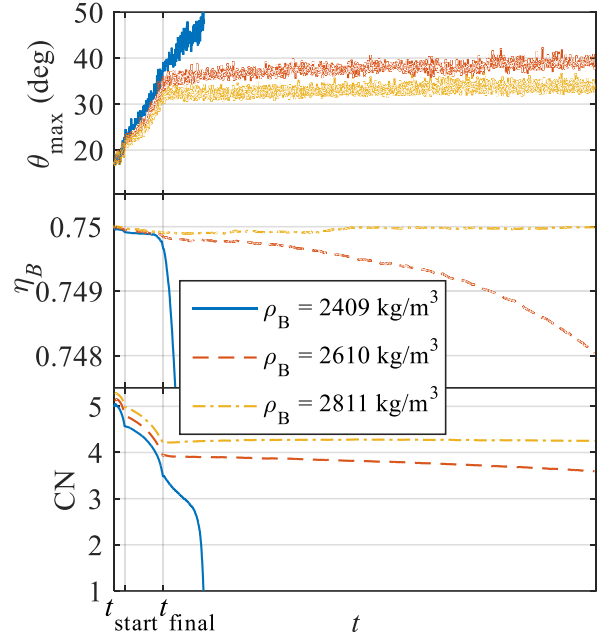

**Figure 22** Evolution of the maximum mobilized friction angle, the internal packing efficiency, and the coordination number for the PP2 model for 3 values of bulk density, $\rho_B$. Path No. 2 is applied in this case. Gradual increases in the maximum mobilized friction angle or gradual decreases in the internal packing efficiency and coordination number after $t_{\text{final}}$ are indicators of "creep".

Only those cases where the body can maintain its shape without further creep motion can be regarded as geo-statically stable cases. As shown in Fig. 22, the case of $\rho_B = 2811$ kg/m$^3$ presents the typical behavior of a stable case, in which the internal packing efficiency and the coordination number will increase slightly after stopping rotational acceleration to offset the dilation caused by the spin-up process, but then reach constant values. The unstable behavior caused by creep deformation is also considered in the process of determining the third critical spin period, $T_{c,3}$ (see Section 3.3.3). We take the spin period where the body can be exactly stable without further creep motion as its third critical spin period.

The above analyses reveal that the third critical spin period and the associated angle of friction are independent of the applied spin-up path. Therefore, the third critical spin limit obtained in a non-real-time spin-up simulation is likely to be the actual spin limit of the corresponding structure spun-up by the YORP thermal effect. This study provides a possible criterion to determine the actual spin limit.

4.4 Effect of friction resistance

The friction resistance between particles has a significant impact on the material shear strength for a granular assembly, which can also affect the spin limit of rubble-pile bodies (e.g., Holsapple, 2001; Sánchez and Scheeres, 2012). Materials of different friction resistance can be achieved by adjusting the shape parameter $\beta$ and the coefficient of static friction $\mu_S$. We investigated three sets of material parameters for this study, namely, $\beta = 0.3$ and $\mu_S = 0.5$, $\beta = 0.5$ and $\mu_S = 1.0$ (the nominal material parameters), and $\beta = 0.6$ and $\mu_S = 1.3$. The HCP configuration and the PP2 configuration are again taken as representative cases.

Our simulations show that the evolution of state variables for different material parameters follows the same trend and the same failure mode as the nominal case, but with different critical spin limits. Using the same procedure introduced in Section 4.2, the third critical bulk densities and the corresponding prefactors $\kappa_3$ for these cases were calculated, and are presented in Table 7. As expected, increases in the shear strength produce a lower critical bulk density. This implies that the surface friction and rotational resistance are also a source of creep stability.

**Table 7** Critical bulk density for HCP and PP2 models with different material parameters. See Sections 2.3.3 and 4.2 for the symbol definitions.

| Rubble-pile Model | $\beta = 0.3, \mu_S = 0.5$ | | | $\beta = 0.5, \mu_S = 1.0$ | | | $\beta = 0.6, \mu_S = 1.3$ | | |
|---|---|---|---|---|---|---|---|---|---|
| | $\rho_B^{c,3}$ (kg/m³) | $\kappa_3$ | $\phi$ (deg) | $\rho_B^{c,3}$ (kg/m³) | $\kappa_3$ | $\phi$ (deg) | $\rho_B^{c,3}$ (kg/m³) | $\kappa_3$ | $\phi$ (deg) |
| HCP | 2430 | 0.937 | 42 | 2383 | 0.946 | 43 | 2369 | 0.949 | 43 |
| PP2 | 2952 | 0.850 | 35 | 2662 | 0.895 | 38 | 2612 | 0.904 | 39 |

Comparing the differences in the friction angles and the critical bulk densities caused by the material parameters in different models, it seems that the PP2 model is more sensitive to variations in the material parameters. As discussed in Section 4.1.1, the shear resistance of HCP configurations mainly arises from geometric interlocking and partly arises from the friction resistance. Compared to the other random packing structures, the ma-terial parameters have a smaller impact on this HCP structure. Even without including the interparticle friction, the friction angle of the HCP model could also be close to 40° (Walsh et al., 2012).

**5. Discussion**

The simulation results presented in Section 4 show the complexity of the effect of configuration, and the dependence of the critical spin limits on bulk density, spin-up path, and material parameters. This section summarizes the implications of the measured critical spin limits and discusses the possible physical properties of the Didymos primary and the formation mechanism of the secondary.

5.1 Implications of the critical spin limits

The three critical spin limits are designed to mark the starting points of reshaping behavior, internal deformation, and structural yielding, as introduced in Section 3. $T_{c,1}$ and $T_{c,2}$ can be directly determined from the evolution curves of the axis ratios and the internal packing efficiency of a simulation, respectively. Although there is no apparent relation between $T_{c,1}$ and $T_{c,2}$, their relative magnitude seems to be associated with the failure mode for each configuration. As listed in Table 3, $T_{c,2}$ is greater than $T_{c,1}$ for configurations with an interior failure mode (i.e., RCP, PP1, and PP2), while $T_{c,2}$ is less than $T_{c,1}$ for configurations with a surface-shedding failure mode (i.e., HCP and PPC). The rule holds true regardless of the bulk density, the spin-up path, and the material parameters (see Fig. 19 and Table 6). The RCPC model is an exception because the surface shedding also trigger changes in its internal packing efficiency, as discussed in Section 4.1.3. In this way, the failure mode of a spinning rubble pile may be inferred by comparing the value of $T_{c,1}$ and $T_{c,2}$.

By definition, the third critical spin period $T_{c,3}$ is the actual spin period at which the body can exactly maintain a stable shape in its current arrangement without further creep deformation regardless of spin-up path. Although the value of $T_{c,3}$ is determined at the spin state where a local region in the rubble-pile structure is about to yield, $T_{c,3}$ also gives the global failure condition for the five random models (except for the HCP model, whose surface fails locally first, so landslides can reduce the surface slope, leading to a stable structure, as discussed in Harris et al., 2009) since the starting point of failure in these rubble-pile models always occurs locally and internally (Hirabayashi et al., 2015) and the initiation of localized yielding can spread and lead to global disruption.[13] Therefore, given that the behavior of deformation or reshaping is just the mechanical response of granular material to structural failure, $T_{c,3}$ is always greater than $T_{c,1}$ and $T_{c,2}$.

5.1.1 Spin limit analyses: comparison with continuum theory

Associated with $T_{c,3}$, we proposed a method in Section 3.3.3 to determine the angle of friction for a spinning self-gravitating granular assembly. As shown in Table 3, a higher friction angle will lead to a lower $T_{c,3}$, implying a stronger structure. Several analytical methods based on continuum theory have been developed to reveal the relation between the spin limit of a self-gravitating body and the material friction angle, which can be divided into two categories—the global-volume-averaging method (e.g., Holsapple 2007; Sharma et al., 2009) and the method to analyze the local stress distribution in a body (e.g., Hirabayashi, 2015). As discussed in Hirabayashi et al. (2015), the global-volume-averaging method can be treated as an upper bound condition on the critical spin rate (i.e., the *theoretical upper spin limit*) where the structure fails globally, while the second method can serve as a lower bound condition on the critical spin rate where only a local region reaches its failure condition (i.e., the *theoretical lower spin limit*). And the actual critical spin rate is between the two spin limits (see Section 2.2 in Hirabayashi et al., 2015 for details).

Figure 23 presents the scaled third critical spin limit, $\Omega_{c,3}/\sqrt{4\pi G \rho_B/3}$, (i.e., the spin-limit proportionality constant $\kappa_3$ listed in Table 4), for our six models with the nominal material parameters compared to the theoretical upper spin limit derived by Holsapple (2007) and the theoretical lower spin limit given in Hirabayashi (2015) for a cohesionless homogeneous spherical body.[14] As shown in Fig. 23, the spin limit of the RCP model is well described by the continuum theory, while the other three homogeneous (i.e., HCP, PP1, and PP2) models plot above the theoretical upper spin limit. The differences between our results and the analytical theory are consistent with our findings that the spin limit is sensitive to the particle arrangement in the discrete case, which cannot be well captured in a continuum method. Although the behavior that the macroscopic heterogeneous cases (i.e., RCPC and PPC) that resist interior failure can maintain the shape at even higher spin rates is in agreement with the continuum theory (e.g., Hirabayashi et al., 2015), it should be noted that the microscopic features of a granular assembly are extremely heterogeneous even though its macroscopic properties are uniform. Compared to other configurations, the microscopic heterogeneity of the RCP configuration is relatively low due to its monodispersity and random arrangement (the crystallized structure of the HCP configuration results in high heterogeneity between the surface and internal regions), which makes its behavior a better match to the continuum theory. [15]

---

[13] Note that we test the creep stability of the simulated rubble pile by forcing it to stay at the spin period of $T_{final}$ (see Section 3.3.3). When a local region in the internal structure yields, the particles inside this region become mobilized and transfer momentum to the neighboring particles. Due to the restriction of particle resolution and spherical shape in a DEM simulation, the dissipative mechanism of particle contacts and collisions cannot balance the shear stress in these simulated cases, resulting in global creep deformation. However, in nature, it is possible that the local region where the body is about to yield can be rearranged to a configuration with higher shear strength, and would not yield to global creep. This can be investigated further with higher-resolution simulations and by replacing the spherical particles with polyhedra once that capability is available for these kinds of DEM simulations.

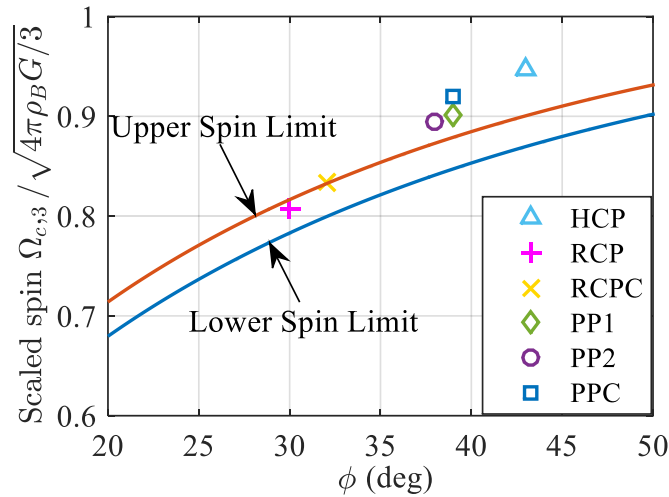

---

[14] Both theoretical spin limits are derived using the Drucker-Prager yield criterion (see Footnote 8). The lower spin limit is given by Equation (17) in Hirabayashi (2015), which marks the critical spin rate for a cohesionless homogeneous spherical body where the central region begins to fail.

[15] Note that previous SSDEM simulations on slightly size-heterogeneous packings (e.g., Sánchez and Scheeres, 2012, where particle sizes range from 8 m to 9 m; Hirabayashi et al., 2015, where the size ratio of the largest particle to the smallest one is ~1.4) also showed consistent spin limits and failure behaviors of rubble-pile bodies with the continuum theory, which are similar to our RCP cases.

**Figure 23** The limits for the spin of self-gravitating bodies: the red curve denotes the theoretical upper spin limit derived by Holsapple (2007) and the blue curve denotes the theoretical lower spin limit given in Hirabayashi (2015) for a perfectly spherical body, and the symbols represent the results of our simulations.

Nonetheless, it is difficult to apply the analytical theory derived for a perfectly spherical body to the detailed radar shape model of the Didymos primary, which may also account in part for the differences in Fig. 23. A study on the spin limit of rubble-pile ellipsoids with various shapes will be the subject of future work.

5.2 Possible physical properties of the Didymos primary

Apart from its spectral properties and radar images, little is known about the physical properties of the Didymos primary, especially its internal structure and particle size distribution. Assuming the asteroid consists of cohesionless granular materials, some constraints can be put on its configuration based on the current observed characteristics.

The first characteristic is the ability to produce the secondary.[16] Since the mass ratio between Didymos' secondary and primary is believed to be less than 1%, the configuration and failure mode of the primary should be similar to the parent body of the system. Therefore, the primary may also be able to produce a secondary at a certain spin state. As presented in Section 4.1, the failure mode of a rubble pile mainly depends on its configuration. Among various formation mechanisms of binary asteroids, the landslide and material shedding mechanism can well explain the observed features of those binary asteroids which have similar properties to the Didymos system (Walsh and Jacobson, 2015). Previous analyses using continuum theory suggest that a self-gravitating homogeneous rubble pile will reshape and deform internally to inhibit surface failure and mass shedding behavior (Holsapple, 2010; Hirabayashi and Scheeres, 2015). In the six configurations explored in this study using SSDEM, only the RCP model exhibits this pure-deformation behavior, while the others (even those macroscopic homogeneous models) all have the ability to produce orbiting material in the equatorial plane through surface shedding. The inconsistency is due to the fact that the surface material motion and shedding in a rubble pile are stimulated by the microscopic heterogeneity that is not included in the continuum methods.

---

[16] Although there are other mechanisms to form a binary system, such as reaccumulation after catastrophic disruption (Michel et al., 2001; Durda et al., 2004), the rotational-fission formation mechanism induced by the YORP effect is a better match to the properties of the Didymos system (Pravec, 2010; Walsh and Jacobson, 2015), so it is preferred in this study.

The second characteristic is the ability to be stable at the observed spin rate. Although current observations cannot rule out the possibility that the primary is still in the process of shedding mass or deforming, it is useful to compare the creep stability of different configurations. As indicated in Table 3, the particle size and distribution of an aggregate have a significant impact on its shear strength. The HCP structure attains the highest shear strength among the six configurations due to the geometric interlocking effect. Increasing the polydispersity and the packing efficiency can also enhance the stability of a spinning granular assembly. Additionally, the existence of a core with higher shear resistance or higher density will increase the critical spin rate of an aggregate, but via different mechanisms. For random uniform structures (e.g., RCP and PP1/2), the central region is subject to the highest shear stresses. A higher friction angle can keep the internal core stable at a higher spin rate by enhancing the shear strength of the internal structure and postponing the failure point to the moment when the external shell fails, while the denser core exerts its effects through imposing higher gravitational pull and easing the shear stresses in the internal core.

As discussed in Section 4.2, the spin limit of a rubble pile is associated with its bulk density. Since the third critical spin period can represent the actual spin period regardless of the spin-up path (see Section 4.3), we take the third critical bulk density as the actual critical bulk density in the following analyses. As indicated in Table 4, only four models (HCP, PP1/2, and PPC) can be stable within the uncertainty of the observed bulk density, i.e., 2100 kg/m$^3$ ± 30%. The critical grain densities are 3370 kg/m$^3$, 3764 kg/m$^3$, 4056 kg/m$^3$ and (3360 and 5040) kg/m$^3$ for the HCP, PP1, PP2, and PPC models, respectively. Given its spectral properties, Didymos is classified as an S-type asteroid (de León et al., 2010; Pravec et al., 2012). The reflection spectrum of S-type asteroids shows similar characteristics to that of the ordinary chondrites (Gaffey et al., 1993), which was confirmed later in the analyses of the sample of the S-type asteroid Itokawa obtained by the Hayabusa mission (Yurimoto et al., 2011). Based on the grain densities and porosities of various ordinary chondrites, some studies (e.g., Flynn et al., 1999; Britt et al., 2002; Carry, 2012) suggest the S-type asteroids have grain densities of between approximately 3100 and 3800 kg/m$^3$ (see Table I in Flynn et al., 1999 and Table 2 in Britt et al., 2002). This narrows the possible configurations to the HCP model and the highly heterogeneous PP1 model.

Polydisperse configurations are most reasonable according to the current known regolith particle size distribution on asteroids (e.g., Thomas et al., 2002; Michikami et al., 2010; Sierks et al., 2011), which also satisfy the requirements discussed above (i.e., the abilities to generate a secondary and be stable with reasonable bulk density). Given that the creep stability of a rubble pile can be improved with a higher internal bulk density, the PPC configuration with a smaller grain density ratio is still reasonable. Through the AIDA mission, the information about the morphology, porosity, and density distribution of the Didymos primary could constrain the grain size and density distribution inside the asteroid (Michel et al., 2016); based on this, a more precise discrete model could be built and its failure behavior and condition could be revealed using the same procedure in this study.

In general, with a material friction angle ~39°, the Didymos primary needs a significantly high density and high polydispersity to maintain its current shape. Although adjusting the material parameters can improve the shear strength of its structure, higher friction angles seem to be unachievable for a cohesionless asteroid soil. That is to say, some cohesion would be required if a lower bulk density for the asteroid is expected. Besides, the tidal pull of the secondary on the primary will disturb its structural stability (Sharma, 2015, 2016), implying the necessity of even higher bulk density or the presence of cohesion. Investigation of cohesion and the effect of the secondary is left for a follow-up study (Zhang et al. 2017, in preparation).

5.3 A plausible binary formation mechanism

This study reveals that surface shedding caused by rotational instability would occur in a macroscopic uniform granular asteroid. Assuming the Didymos parent body has a structure similar to the PP1 configuration, the internal region begins to deform when exceeding the failure limit. The poles of the aggregate push inwards toward the center, leading to a more oblate shape. Soon afterward, some surface particles can no longer be constrained by their neighbors and trigger a landslide. These mobilized materials flow to the equatorial region, and would be shed and leave the asteroid surface, which could re-accumulate in orbit to form the secondary. The formation of the equatorial ridge is the result of internal deformation and landslides. The mechanism of deformation and mass shedding is a combination of the previous proposed theories (i.e., Hirabayashi and Scheeres, 2015; Walsh et al., 2008), which can well explain the currently understood characteristics of the Didymos system.

As discussed above, we found conditions under which a rubble pile with the Didymos primary's radar shape could reproduce the current observed characteristics. How the Didymos primary arrived at this shape is not simulated in this work. But, we can speculate that rotational stresses drove it to its present configuration, and may have led to the formation of the secondary. If successful, the AIDA mission should provide insight into the dynamical history of this intriguing system. In particular, images may show the traces of landslides at latitudes between the pole and the equator. Also, the secondary may be a rubble pile if the system is formed by the mechanism proposed above, which can be checked thanks to the proposed low-frequency radar instrument of the AIM spacecraft (Michel et al., 2016).

## 6. Conclusions

In this paper we described numerical tests to investigate the creep stability and possible physical properties of the Didymos primary by representing it as a granular assembly. From a granular mechanics perspective, the shear strength of an aggregate strongly depends on the arrangement and size distribution of constituent particles and interparticle friction, which was systematically studied in this paper. We found that: (1) for assemblies with crystal structure, geometric interlocking caused by the arrangement of particles is the main source of shear strength; (2) for random packings, increased size heterogeneity can generate an assembly with a higher packing efficiency and higher shear strength; (3) higher interparticle friction can keep a spinning rubble pile stable at a higher spin rate; and (4) the failure mode of a rubble pile mainly depends on its configuration (i.e., the arrangement, size distribution, and density distribution of particles) regardless of the bulk density and interparticle friction. Our results also confirmed that the spin limit of a self-gravitating aggregate has a square-root dependency on its bulk density. With a highly size-heterogeneous configuration, the Didymos primary rubble-pile model can maintain its shape at the current observed spin rate within the uncertainty of the observed bulk density.

Furthermore, we proposed a new method to assess the spin limit of a self-gravitating rubble pile through numerical simulations. Three critical spin limits were devised to indicate the critical states triggered by variations in external shape and surface shedding, internal structural deformation, and shear yielding, respectively. The failure information (i.e., the failure spin and failure mode) and strength of a rubble pile can all be inferred from the three critical spin limits. If the first critical spin limit is larger than the second one, the corresponding rubble-pile configuration begins to fail internally; otherwise, the rubble pile exhibits surface-shedding failure behavior. The third critical spin limit is determined by evaluating the Mohr-Coulomb yield criterion, which also provides a method to assess the macroscopic material friction angle for a spinning self-gravitating aggregate. Since the third critical spin limit is independent of the spin-up path, this study provides a possible criterion to determine the actual spin limit of a rubble pile subject to realistic slow rotational acceleration.

The present study was intended to systematically assess the creep stability of a spinning self-gravitating rubble pile with a realistic irregular asteroid shape from a granular mechanics perspective. Future studies could investigate a greater range of configurations (such as a size distribution weighted by distance from the center of mass), explore the effect of cohesion, and extend the method proposed in this study to other asteroids with different shapes and physical properties. Regarding the proposed AIDA mission, it is important to understand the effect of the secondary on the structural stability of the primary and the overall orbital stability, a study that is currently underway (Zhang et al., 2017, in preparation).

**Appendix: The variation in packing efficiency**

Considering two particles of radius $\bar{r}$ in contact with each other (Fig. A), the variation in the volume of the particle's Voronoi cell caused by this contact can be approximated by

$$\Delta V = \pi \bar{r}^2 x/2. \tag{A.1}$$

Suppose the original packing efficiency of this particle is $\eta_p$. With an overlap of $x$, the variation in its packing efficiency after removing this contact is

$$\Delta \eta_p = \eta_p - \frac{4\pi \bar{r}^3/3}{\Delta V + 4\pi \bar{r}^3/(3\eta_p)}. \tag{A.2}$$

Here, the second quantity in the denominator is the volume of the particle's Voronoi cell, in an average sense. For a typical original packing efficiency of 0.6–0.8 and $x$ of 1% $\bar{r}$, $\Delta \eta_p$ is on the order of ~0.001. Since the bulk packing efficiency for a rubble pile is calculated by an averaging method, this magnitude of $\Delta \eta_p$ can also be applied to granular systems with size distributions.

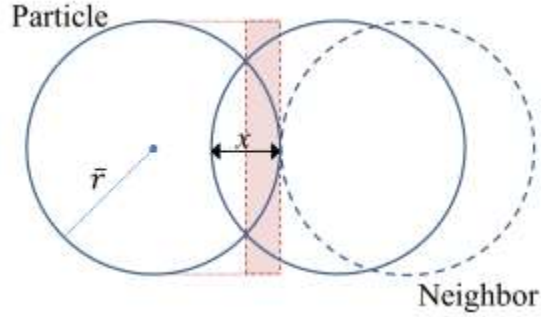

**Figure A** A particle of radius $\bar{r}$ in overlap with one of its neighbors. The quantity $x$ is the scalar distance between the surfaces of the two particles along the line that connects their centers. Note that $x$ is exaggerated to illustrate the method; $x$ typically does not exceed 1% $\bar{r}$. The dashed rectangle demonstrates the reduction in the volume of the particle's Voronoi cell caused by this overlap.

**Acknowledgments**

This work was supported in part by NASA grant NNX15AH90G awarded by the Solar System Workings program. Simulations were carried out at the University of Maryland on the yorp cluster administered by the Department of Astronomy and the deepthought and deepthought2 supercomputing clusters administered by the Division of Informational Technology. Y.Z. acknowledges support from the China scholarship Council. Y.Z. and J.L. acknowledge support from the National Natural Science Foundation of China (NO. 11572166). P.M. acknowledges support from the European Space Agency (ESA RFP/NC/IPL-PTE/LF/as/221.2015, AIM Advisory Study).